\documentclass{jfm}

\usepackage{graphicx}
\usepackage{newtxtext}
\usepackage{newtxmath}
\usepackage{mathtools}
\usepackage{amsmath}
\usepackage{bm}

\usepackage{natbib}
\usepackage{hyperref}
\hypersetup{
    colorlinks = true,
    urlcolor   = blue,
    citecolor  = black,
}

\usepackage{multirow}
\usepackage{subcaption}
\usepackage{xcolor} 
\usepackage{epstopdf, epsfig}
\graphicspath{{figures/}}
\usepackage{cleveref} 

\newcommand{\RomanNumeralCaps}[1]
\linenumbers
\graphicspath{{figures/}}
\usepackage{cleveref} 

\shorttitle{Stratified inclined duct: DNS}
\shortauthor{L. Zhu, A. Atoufi, and others}

\newcommand{\hg}{{\bf \hat{g}}}

\title{Stratified inclined duct: direct numerical simulations}

\author{Lu Zhu\aff{1}, Amir Atoufi\aff{1}
	\corresp{\email{aa2295@cam.ac.uk}}, Adrien Lefauve\aff{1}, John R. Taylor\aff{1}, 
	Rich R. Kerswell\aff{1},
	Stuart B. Dalziel\aff{1},  Gregory. A. Lawrence\aff{2}, \and P. F. Linden\aff{1}}

\affiliation{\aff{1}Department of Applied Mathematics and Theoretical Physics, Centre for Mathematical Sciences, Wilberforce Road, Cambridge CB3 0WA, UK
	\aff{2} Department of Civil Engineering, University of British Columbia, Vancouver, BC V6T 1Z4, Canada}

\begin{document}
	
	\maketitle

	\begin{abstract}
		 The stratified inclined duct (SID) experiment consists of a zero-net-volume exchange flow in a long tilted rectangular duct, which allows the study of realistic stratified shear flows with  sustained internal forcing.
		 We present the first three-dimensional direct numerical simulations (DNS) of  SID to explore the transitions between increasingly turbulent flow regimes first described by Meyer \& Linden (\textit{J. Fluid Mech.} \textbf{753}, 242-253, 2014). We develop a numerical set-up that faithfully reproduces the experiments and sustains the flow for arbitrarily long times at minimal computational cost.
	    We recover the four qualitative flow regimes found experimentally in the same regions of parameter space: laminar flow, waves, intermittent turbulence, and fully-developed turbulence. We find good qualitative and quantitative agreement between DNS and experiments and highlight the added value of DNS to complement experimental diagnostics and increase our understanding of the transition to turbulence, both temporally (laminar/turbulent cycles) and parametrically (as the tilt angle of the duct and the Reynolds number are increased).  
	    These results demonstrate that numerical studies of SID -- and deeper integration between simulations and experiments -- have the potential to lead to a better understanding of stratified turbulence in environmental flows.

	\end{abstract}
	

\section{Introduction}\label{sec:intro}
	
Large-scale fluid motions in the ocean are almost always stably-stratified in density due to differences in temperature and/or salinity at different depths. The transport of momentum and mass (temperature, salinity, and other solutes) by turbulence plays an important role in setting the large-scale structure and circulation of the ocean, with implications for the global climate. Consequently, the influence of stable stratification on turbulence and the resulting mixing has attracted much attention \citep{linden1979mixing,riley2000fluid,gregg_mixing_2018,colm2020open,dauxois_confronting_2021,caulfield2021layering}.

Sustained stratified shear-driven flows are
a particularly interesting and relevant class of flows to study this problem, since turbulence is produced internally within the flow by drawing energy from the background shear, and because turbulence persists
over sufficiently long periods of time to allow for a statistically-steady dissipative equilibrium. 
The stratified inclined duct (SID) experiment was developed
to study these flows in a controlled laboratory environment \citep{meyer2014stratified}. It establishes a two-layer exchange flow through a long, rectangular, and slightly inclined duct connecting two large reservoirs containing fluids of different densities. The SID experiment revealed that the flow regime within the duct could be tuned by adjusting the tilt angle $\theta$ of the duct with respect to the horizontal, and/or the Reynolds number $\mathrm{Re}$ based on the initial density difference and the height of the duct. The flow regimes are (ordered by increasing $\theta \mathrm{Re}$):  laminar two-layer flow, interfacial waves, intermittent turbulence with increased interfacial mixing, and eventually full turbulence with significant mixing. Much insight has already been gained through  experimental studies of these regimes and of their transitions \citep{meyer2014stratified,lefauve2020buoyancy,lefauve_experimental1_2022}, of their energetics and mixing properties \citep{lefauve2019regime,lefauve_experimental2_2022}, and of their respective coherent structures  \citep{lefauve2018structure,jiang_evolution_2022}.

Despite vast technological improvements yielding unprecedented time-resolved, volumetric velocity and density data \citep{partridge_versatile_2019}, experimental limitations remain. The SID experimental data do not yet cover the full length of the duct, do not yet achieve the spatial resolution required to fully quantify energy dissipation and mixing, and are not yet as instantaneous and accurate as we would ideally like (due to their reconstruction of volume by successive scanning of planes). In this paper, we present the first direct numerical simulations (DNS) of SID to help overcome these limitations and integrate experiments and simulations more deeply. 

Previous DNS of stratified shear flows have considered more idealised problems, typically without any forcing to sustain the flow \cite[e.g.][]{salehipour_self_2018,watanabe_hairpin_2019}. The boundary conditions are usually idealised too, being typically periodic for velocity and density in the streamwise and spanwise directions. By contrast, experiments have revealed that the specific `natural' forcing mechanisms in SID flows (a streamwise hydrostatic pressure gradient and the tilt angle $\theta$) and the lack of periodicity in the streamwise direction (i.e. the presence of reservoirs) are essential features that need to be  modelled accurately in order to understand this canonical flow. For example, these features are thought to be closely linked to the notion of `hydraulic control' of the exchange flow at high enough values of $\theta \mathrm{Re}$, and to the ensuing transition to turbulence \citep{meyer2014stratified,lefauve2019regime,lefauve2020buoyancy}. The no-slip boundary conditions at the duct walls have also been shown to be important to the structures of instabilities  \citep{lefauve2018waves,ducimetiere2021effects}. 

In  \S\ref{sec:method}  we explain how we overcome the challenges of developing faithful DNS of SID. In particular, we discuss how we modelled the reservoirs with minimal computational cost, and how we handled technically challenging boundary conditions. In \S\ref{sec:valid} we  validate this new DNS methodology by comparing different reservoir geometries and forcing with fully-resolved computations that capture the reservoirs explicitly.
In \S\ref{sec:transition}, we  describe the flow regimes and compare the DNS with experimental data, first from regime diagrams (i.e. the map of the observed qualitative flow regimes in the two-dimensional parameter space $\theta-\mathrm{Re}$), and then from shadowgraph visualisations of the density interfaces. 
Then in \S\ref{sec:addedvalue} we describe further quantitative diagnostics from our DNS, generally inaccessible to experiments, and highlight their added value. These include the gradient Richardson number,
{the turbulent kinetic energy and pressure fields along the entire length of the duct, and the turbulent energy fluxes.} Finally, in \S\ref{sec:conclu} we conclude by summarising our results, open questions and  outlook.

	\section{Methodology}\label{sec:method}

	\subsection{Governing equations} \label{sec:gov-eqs}

\begin{figure}
		\centering		
		\includegraphics[width=.9\linewidth, trim=0mm 0mm 0mm 0mm, clip]{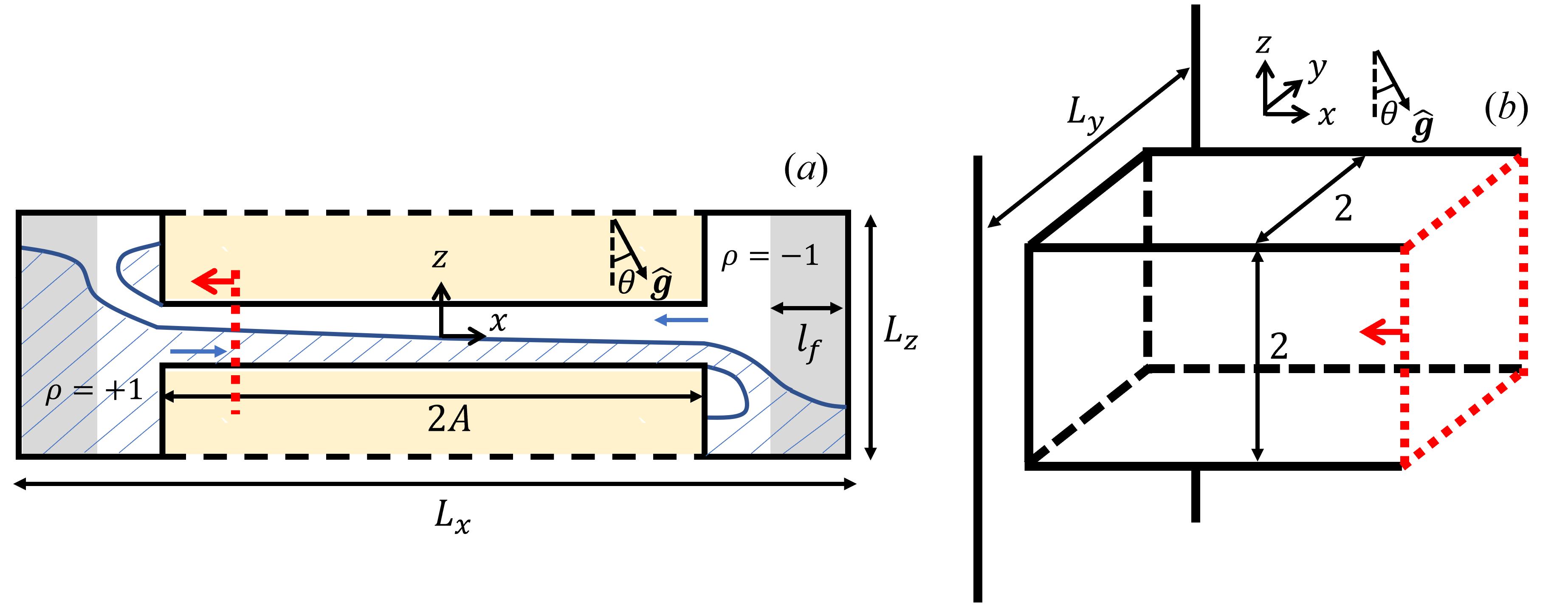}
		\\
		\includegraphics[width=.9\linewidth, trim=5mm 0mm 0mm 0mm, clip]{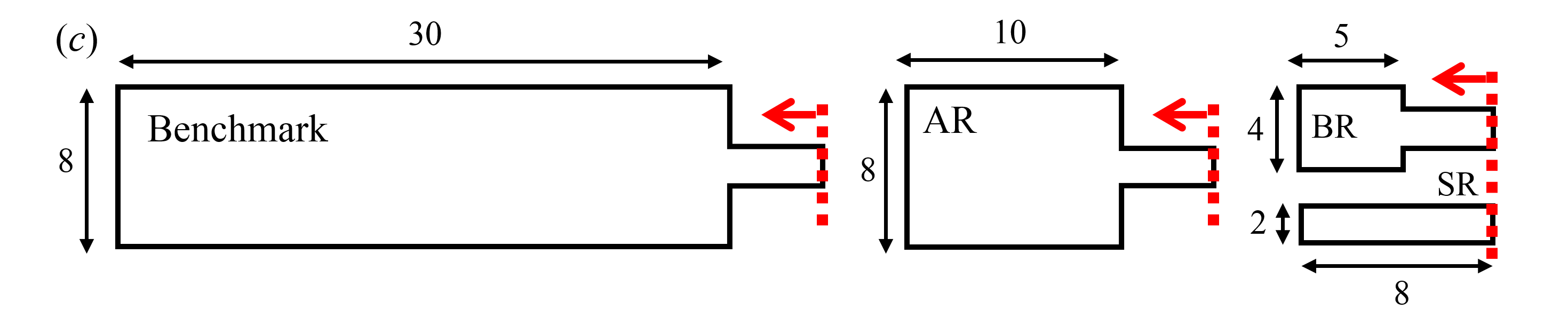}
		\caption{Schematics of SID geometry in non-dimensional units. (a) Overview of the rectangular simulation domain of dimensions $L_x,L_y,L_z$ within which immersed boundaries create a square duct of dimensions $2A\times2\times2$. (b) Detail of the duct geometry and coordinate system. (c) Shape of the different reservoirs considered in this paper (Bench., AR, BR, and SR), with the total domain length $L_x=2(A+L^r_x)$. All numerical parameters are summarised in~\cref{tab:simul_overview}. }
		\label{fig:geom}
	\end{figure}

   Our simulation geometry in non-dimensional units is shown in figure~\ref{fig:geom}(a,b). 
   It replicates the experimental geometry (see, e.g. figure 1 of \cite{lefauve2019regime} in dimensional units), which consists of a duct of square cross-section with internal height $H$, width $W$ and length $L$ connecting two large reservoirs with fluids at densities $\rho_0\pm \Delta\rho/2$ { (white and blue shaded areas in figure~\ref{fig:geom}(a))}. To match previous experimental studies of SID, we non-dimensionalise all lengths  by the duct half height $H/2$, making the duct non-dimensional length, height and width $2A\times 2B\times 2$, respectively, where $A\equiv L/H$ and $B=W/H$ are the streamwise and spanwise aspect ratios, respectively. 
   We also non-dimensionalise (i) the velocities by the fixed buoyancy velocity scale $\Delta U/2 \equiv \sqrt{g^\prime H}$ (where $g^\prime=g\Delta \rho/\rho_0$ is the reduced gravity and $\rho_0$ is the reference density); (ii) the time by the advective time unit (ATU)  $H/\Delta U$; (iii)  the density variations around the reference $\rho_0$ by $\Delta\rho/2$; and (iv) the pressure by $\rho_0(\Delta U/2)^2$. Note that  the $x$-axis (the streamwise direction) is aligned along the duct, whereas the gravity points downwards at an angle $\theta$ from the $-z$ axis (the vertical direction in the frame of the duct), hence in these duct coordinates $\mathbf{g}=g
  \hg=g(\sin{\theta}, 0, -\cos{\theta})$. 
	
	The resulting non-dimensional governing equations for our DNS are the Navier-Stokes equations under the Boussinesq approximation
	\begin{eqnarray}
		\label{eq:ns_mass}%
		\boldsymbol{\nabla} \cdot \mathbf{u} &=& 0,%
		\\
		\label{eq:ns_mom}%
		\frac{D \mathbf{u}}{D t} &=&
		- \boldsymbol{\nabla}p + \frac{1}{\mathrm{Re}} \nabla^{2}\mathbf{u} 
		+ \mathrm{Ri} \, \rho \, \hg  - \boldsymbol{F}_u,%
		\\
		\label{eq:ns_den}%
		\frac{D \rho}{D t} &=&
		\frac{1}{\mathrm{Re \ Pr}} \nabla^{2}\rho  - F_\rho,%
	\end{eqnarray}
	where the material derivative is $D/Dt \equiv \partial_t+\mathbf{u}\cdot \boldsymbol{\nabla}$, the velocity is
	$\mathbf{u}=(u,v,w)$ is  in the non-dimensional coordinate system $\mathbf{x}=(x,y,z)$ aligned with the duct,  the non-dimensional pressure is $p$, the non-dimensional density variation around the mean is $\rho$ (bounded between $-1$ and $1$). The forcing terms $\boldsymbol{F}_u$ and $F_\rho$ used to maintain the quasi-steady exchange flows will be described in \S\ref{sec:force}. 
	
%
%


	The non-dimensional Reynolds, Richardson, and Prandtl numbers are  related to the dimensional experimental parameters as follows
	\begin{equation}
	    \mathrm{Re}\equiv \frac{\frac{\Delta U}{2}\frac{H}{2}}{\nu} \equiv \frac{\sqrt{g'H}H}{2\nu}, \qquad \mathrm{Ri}\equiv \frac{\frac{g}{\rho_0}\frac{\Delta\rho}{2}\frac{H}{2}}{\Big(\frac{\Delta U}{2}\Big)^2}\equiv \frac{1}{4}, \qquad \mathrm{Pr}\equiv \frac{\nu}{\kappa}
	    \equiv 7,
	\end{equation}
	where $\nu$ is the kinematic viscosity and $\kappa$ is the mass diffusivity. Previous studies of SID showed that the streamwise velocity scales with $\Delta U/2$, motivating this definition of Reynolds number. The Richardson number is always equal to 1/4 due to the definition of $\Delta U$. The Prandtl number in all simulations was set to $\mathrm{Pr}=7$, approximately representative of temperature stratification in water at room temperature. For a given duct and reservoir geometry, there are two remaining free non-dimensional parameters: the tilt angle $\theta$ and the Reynolds number $\mathrm{Re}$ (based on the driving density difference $\Delta\rho$).

	\subsection{Artificial restoring of the exchange flow}	
	\label{sec:force}
	The exchange flow in the duct is driven by the hydrostatic longitudinal pressure gradient and by the longitudinal gravitational acceleration $g\sin\theta$. 
    {In the context of the two-layer flow, the along-duct component of gravity  accelerates the heavier layer rightwards (downhill) and the lighter layer leftwards (uphill). The role of the hydrostatic pressure gradient turns out to be more intricate and will be examined in \S\ref{sec:pressure}.}
	In the experiments the flow inside the duct is sustained over long time periods (typically several hundred advective time units) until the discharged fluids accumulated in the large reservoirs have reached the level of the duct. Simulating such large reservoirs would be prohibitively expensive. In the simulations, we use smaller reservoirs and add \textit{ad hoc} forcing terms
	$\boldsymbol{F}_u, F_\rho$ in the momentum and buoyancy  equations \eqref{eq:ns_mom} and \eqref{eq:ns_den}, respectively,
	
	

	\begin{equation} \label{eq:F_v}
		\boldsymbol{F}_u \equiv  F_u \mathbf{u}  \equiv
		\Big[ \frac{1-\tanh(\frac{2}{\Delta}(x+\frac{L_x-l_{f}}{2}))}{\eta_u}+\frac{1+\tanh(\frac{2}{\Delta}(x-\frac{L_x-l_{f}}{2}))}{\eta_u} \Big] \mathbf{u} ,%
	\end{equation}
	\begin{equation} \label{eq:F_ro}
		F_\rho\equiv \frac{1-\tanh(\frac{2}{\Delta}(x+\frac{L_x-l_{f}}{2}))}{\eta_\rho}(\rho-1)+\frac{1+\tanh(\frac{2}{\Delta}(x-\frac{L_x-l_{f}}{2}))}{\eta_\rho}(\rho+1) ,
	\end{equation}
	where $l_{f}$ is the streamwise length of influence of the forcing, and $\Delta=2  l_f/L_f$ (with a fixed $L_f\equiv 8$) defines 
	the steepness of the transition from the forced to the unforced regions. The density forcing term restores the density of the fluid in the reservoir to the prescribed value (i.e., $\pm 1$), and the momentum forcing term acts to dampen motion in the reservoir. The timescales $\eta_u$ and $\eta_\rho$  control the momentum and density forcing terms, respectively. {Compromise values of these timescales must be found, as large values are too slow to sufficiently damp reservoir motion and restore density, while small values are too fast and overreact, threatening numerical stability.  } 
	These parameters were optimised with the size of the reservoirs in order to minimise their influence on the large-scale flow in the duct compared to the Bench. cases with large reservoirs and without forcing (see \S\ref{sec:res-geom}). 
	{Tests revealed little variation in  the range $l_{f}\in [0.3L_x^r ,0.7L_x^r]$, therefore we set  $l_f=0.5L_x^r$, confining the forcing region  to  half the reservoir (greyed out in figure~\ref{fig:geom}(a)). The timescales $\eta_u$ and $\eta_\rho$ should then be smaller than the times for a discharging flow (with non-dimensional speed 1) to pass through the forcing region, i.e. $\approx l_f$. Practically, we set
	$2.5 \lesssim \eta_u \lesssim 5$ and $0.1 \lesssim \eta_\rho \lesssim 0.5$, depending on $l_f$. } 

	Physically, $\boldsymbol{F}_u$ decelerates the fluid entering the reservoir until it comes to rest, and	
	$F_\rho$ ensures that the density of 
	this fluid matches that of the reservoir before it re-enters the duct.
	This forcing thus effectively mimics the action of infinitely large reservoirs with a finite-sized, computationally-feasible domain.
	
	\subsection{Solver}
	The DNS were performed with the open-source solver Xcompact3D~\citep{Bartholomew2020xcompact3d}, which uses 4th-order and 6th-order compact finite-difference schemes for the first and second spatial derivatives, respectively, and a 3rd-order Adams-Bashforth scheme~\citep{peyret2002spectral,Zhu_XiJNNFM2020} for the time integration with a time {step} $\delta_t=0.001$. {Pressure field is obtained from a conventional Poisson equation based on applying divergence operator on the \eqref{eq:ns_mom}, employing continuity \eqref{eq:ns_mass}. The Poisson equation is then solved numerically using the fast Fourier transform with modified wavenumbers}. For more details about the core of the code (Incompact3D), see \citet{laizet2009high} and \citet{laizet2011incompact3d}, and for the application of Xcompact3D to stratified turbulent flows see \citet{frantz2021high}. We modified Xcompact3D to include the forcing terms $\boldsymbol{F}_u,F_\rho$ discussed above.

	\subsection{Domain and boundary conditions}
	
	The computational domain had dimensions $L_x$, $L_y=2$, and $L_z$ along $x$, $y$, and $z$, respectively (see figure~\ref{fig:geom}(a,b)). 
	On the  boundaries of this  domain, we applied a no-slip  condition for $\mathbf{u}$ and a no-flux condition for $\rho$ as in \citet{laizet2009high}.
	{To represent the duct and reservoir geometry within this computational domain, we applied the immersed boundary method (IBM) in Xcompact3D to the yellow-shaded region in figure~\ref{fig:geom}(a).}

 	\begin{figure}
		\centering		
		\includegraphics[width=.5\linewidth, trim=0mm 0mm 0mm 0mm, clip]{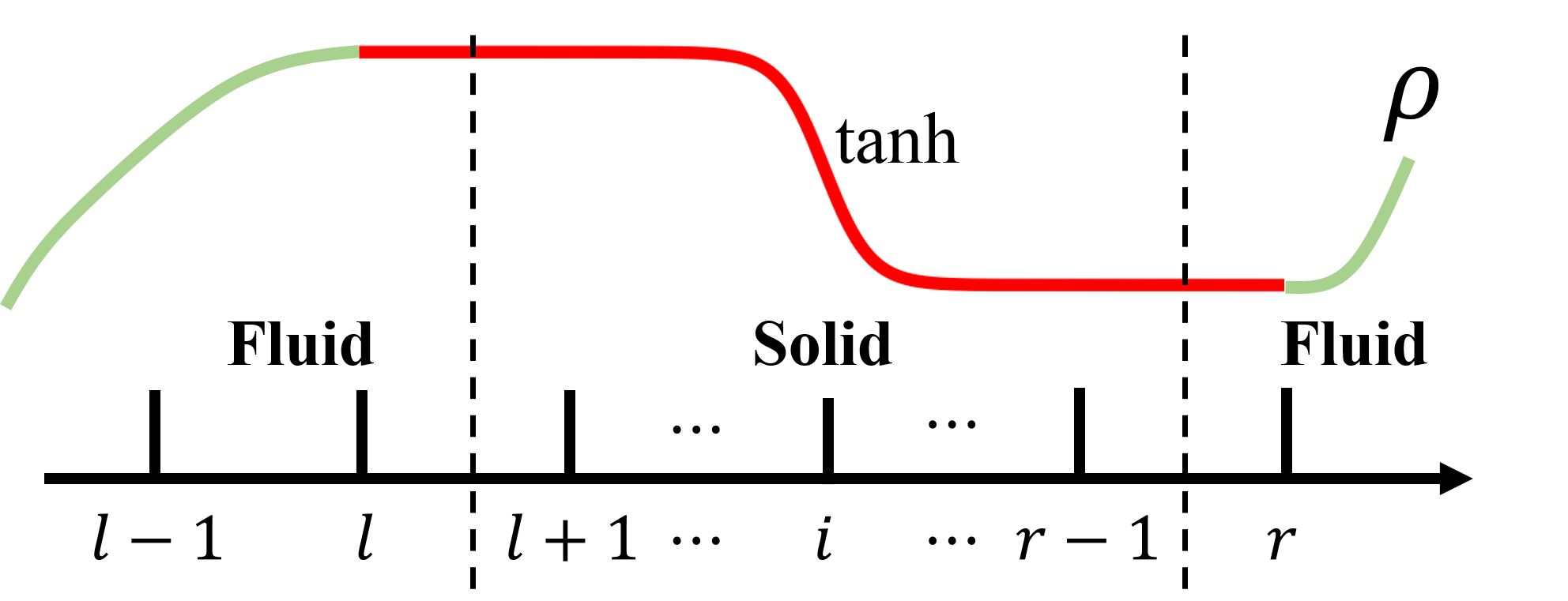}
		\caption{Schematic of the IBM to implement the no-flux boundary condition on density $\rho$ at fluid-solid boundaries {in the $x$ or $z$-direction}. The curved red line represents the fictitious density profile across the solid region.}
		\label{fig:ibm}
	\end{figure}

 {The IBM treatment of $\mathbf{u}$ (no slip) uses a direct forcing method described in \cite{mohd1997combined} and specifically for Xcompact3D in \cite{laizet2009high,Gautier2014}, which imposes $\mathbf{u}=\mathbf{0}$ in the solid regions.
 The pressure $p$ in the solid region is treated by reducing the Poisson equation to a Laplace equation~\citep{laizet2009high}.
 The IBM allows relatively simple implementation of complex geometries in scalable codes such as Xcompact3D that are built upon Cartesian coordinates and rectangular computational domains.    
 }

	
	 
	The IBM treatment of $\rho$ (no flux) required a slightly different approach to minimize the modifications of Xcompact3D and maintain the consistency between $\mathbf{u}$ and $\rho$. A $\tanh$ function was used to reconstruct points inside the solid region
	\begin{equation}
		\rho_i = \frac{1}{2}(\rho_{l}+\rho_{r})+\frac{1}{2}(\rho_{r}+\rho_{l})\tanh\Big[\frac{L_{w}}{\xi_{r}-\xi_{l}}\big(\xi_i-\frac{1}{2}(\xi_{r}+\xi_{l})\big)\Big],
	\end{equation}
	as shown in figure~\ref{fig:ibm}.  Here $\xi$ is the wall-normal coordinate (horizontal or vertical), the subscript $i$ is the index of grid points, $r$ and $l$ are the right and left grid points (in the fluid region), respectively, adjacent to the solid walls. The $\tanh$ function ensures a zero-flux boundary condition at the wall while maintaining a smooth change of the density through the solid region. A similar approach using a polynomial reconstruction has been used to treat the Dirichlet and Neumann boundary conditions in \citet{Gautier2014} and \citet{frantz2021high}. The length scale $L_w=10$ was chosen to ensure a smooth change of density inside the solid region while maintaining exponentially small density flux at SID walls. 

	\subsection{Initial conditions}
	
	All simulations were initialized at $t=0$ with a density $\rho = \tanh(x/L_I)$ (where $L_I=0.1$) at the centre of the duct, simulating `lock exchange' conditions with a sharp but continuous change from densest fluid on the left-hand side to lightest fluid on the right-hand side. 
	{A zero-mean uniform distributed random noise with nondimensional amplitude $\varsigma=0.5$, is applied to the initial velocity $\mathbf{u}_n$. Such random noise is set to break the symmetry of the exchanging flow and initiate instabilities inside the duct. Note that smaller perturbation amplitudes (e.g. $\varsigma=0.005$) can be applied, but we verified it did not influence the main features of the flow (see Supplementary Material S1).}
	
 Shortly after $t>0$, a gravity current formed at the centre of the duct ($x=0$) and propagated in both directions toward the ends of the duct. After a typical duct transit time of order $t\approx A$  (transiting at non-dimensional velocity $\approx1$ over a non-dimensional length $A$), the exchange flow was established.

	\subsection{Parameters, duct  and reservoirs geometries}
	\label{sec:res-geom}
	\begin{table}
		\begin{center}
			\def~{\hphantom{0}}
			{
				\begin{tabular}{lccccccccc}
					$\textrm{Case}$  & $ \mathrm{Re} $  & $ \theta $ (deg.)  & $A$  & $B$ & $L_x^r \times L_y \times L_z$ & $N_x \times N_y \times N_z$ & $l_f$ & $\eta_u$ & $\eta_\rho$ \\[3pt]
					\hline
					\multirow{2}{*}{Bench.}            & 400 & 2  & \multirow{2}{*}{$30$} & \multirow{2}{*}{$1$} & \multirow{2}{*}{$30\times 2\times 8$} & $1621 \times 49 \times 385$ & \multirow{2}{*}{-} & \multirow{2}{*}{-} & \multirow{2}{*}{-}\\
                & 650 & 6  & & & & $1921 \times 61 \times 481$ &  &  & \\
					\hline
					\multirow{5}{*}{AR}   & 400 & 2, 5 & \multirow{5}{*}{$30$} & 
					\multirow{5}{*}{$1$} & \multirow{5}{*}{$10\times2\times8$}& $1081 \times 49 \times 385$ & \multirow{5}{*}{5} & \multirow{5}{*}{5} & \multirow{5}{*}{0.1}\\
					& 650 &  4, 6, 8 & &  &  & $1441 \times 65 \times 481$ & & &\\
					& 800 & 3, 4, 10 & &  &  & $1537 \times 65 \times 577$ & & &\\
					& 1000 & 2, 4 &  & &  & $1729 \times 65 \times 577$ & & &\\
					& 1250 & 1, 3 & &  &  & $1801 \times 81 \times 641$ & & &\\
					\hline
					\multirow{5}{*}{BR} & 400 & 2, 5, 7, 10 & \multirow{5}{*}{$30$} &
					\multirow{5}{*}{$1$} &					\multirow{5}{*}{$5\times2\times 4$} & $961 \times 61 \times 193 $ & \multirow{5}{*}{2.5} & \multirow{5}{*}{2.5} & \multirow{5}{*}{0.1}\\
					& 650 & $\mathbf{2}^{(\mathrm{B2})}$, 4, $\mathbf{5}^{(\mathrm{B5})}$, $\mathbf{6}^{(\mathrm{B6})}$, $\mathbf{8}^{(\mathrm{B8})}$ & & &  & $1081 \times 65 \times 241 $ & & &\\
					& 800 & 7 &  & &  & $1351 \times 65 \times 289 $ & & &\\
					& 1000 & 3, 4, 5, $10$ &  &  &  & $1501 \times 65 \times 281 $ & & &\\
     & 1000 &  $\mathbf{10}^{(\mathrm{B10})}$ &  &  &  & $3001 \times 121 \times 241 $ & & &\\
					& 1250 & 5 &  & &  & $1501 \times 65 \times 289$ & & &\\
					\hline
					\multirow{2}{*}{SR}  & 400 & 2, 5 & \multirow{2}{*}{$30$}& \multirow{2}{*}{$1$} & \multirow{2}{*}{$10 \times 2\times 2$} & $1081 \times 65 \times 121$ & \multirow{2}{*}{8} & \multirow{2}{*}{5} & \multirow{2}{*}{0.5}\\
					& 650 & $\mathbf{6}^{(\mathrm{S6})}$, $\mathbf{8}^{(\mathrm{S8})}$ &  & &  & $1201 \times 65 \times 121$ &  & &  \\				
			 	    \hline
					BRw & 650 &  $\mathbf{3}^{(\mathrm{W3})}$, $\mathbf{5}^{(\mathrm{W5})}$  & $44$ & $2$ & $10\times 4\times 4$ & $1601 \times 121 \times 241$ & 8 & $5$ & 0.5\\	
				    \hline
				\end{tabular}
				\caption{Summary of the DNS, from left to right: reservoir geometry (case) as shown in figure~\ref{fig:geom}(c); Reynolds number; tilt angle; duct streamwise aspect ratio; duct spanwise aspect ratio; reservoir size; grid size of the entire computational domain; and forcing parameters. Bold font and superscripts denote the most used DNS.}\label{tab:simul_overview}
				}
		\end{center}
	\end{table}

    In order to investigate the various flow regimes we varied the Reynolds number $\mathrm{Re}$ in the range $400-1250$ and the duct tilt angle $\theta$ in the range $1-10^\circ$. As mentioned above the Prandtl number $Pr=7$ and  Richardson number $\mathrm{Ri}=1/4$ were fixed. The suite of DNS is summarised in  \cref{tab:simul_overview}.
	 
	Most DNS were run with a long duct of streamwise and spanwise aspect ratios $A=30$ and $B=1$, respectively, for direct comparison with the experiments  in \citet{lefauve2020buoyancy} (the `mini SID Temperature' dataset abbreviated `mSIDT').  However, a couple of DNS (cases `BRw' in \cref{tab:simul_overview}) were run in a longer and wider duct at $A=44$, $B=2$  to compare with a new experimental set-up.
	
	{To validate the performance of our forcing to sustain a realistic exchange flow, we ran a benchmark DNS (`Bench.') without forcing ($F_u=F_\rho=0$) but with large reservoirs ($L_x^r\times L_z=30\times 8$). }
	This benchmark had a combined reservoir volume of four times that of the duct ($60\times8/(60 \times 2)=4$), {which is sufficient for our validation but still much smaller than the experiments (volume ratio $\approx 30$). We will show in \S\ref{sec:valid}, that the different reservoirs do not seem to influence the flow statistics within SID. This is expected from the knowledge that the flow in SID is hydraulically controlled \citep{meyer2014stratified}, i.e. that information from the reservoirs cannot travel into the duct because of `control' regions at the inlet and outlet, where the convective flow speed is faster than interfacial waves \citep{lawrence1990hydraulics}. This conveniently ensures that different reservoir geometries and conditions do not influence the flow within the duct, as long as unmixed and quiescent fluid are available at either end of the duct. }
	 
	 All the other DNS had non-zero forcing and smaller, more computationally affordable reservoirs. To test the impact of reservoir size, we used the three following reservoirs sketched in figure~\ref{fig:geom}(c): the A-reservoir (`AR') of dimensions $L_x^r\times L_z=10\times 8$, which is a third of the length of the Bench but equally tall; the  B-reservoir (`BR') $L_x^r\times\ L_z=5\times 4$, which is half the length and half the height of the A-reservoir; and finally, the smallest S-reservoir (`SR') $L_x^r\times L_z=10\times 2$.  Note that (unlike the experiments) almost all reservoirs have the same spanwise width as the duct $L_y=2$. The only exception is `BRw', which has $L_y=4$ and $L_x^r\times L_z=10\times 2$. The set of forcing parameters $(l_f, \nu_u,\nu_\rho)$ for $\boldsymbol{F}_u,F_\rho$ that we found to have minimal impacts on the duct for each case are also listed in \cref{tab:simul_overview}.

	The bold font for the seven cases at $\mathrm{Re}=650$ and $1000$ highlight the DNS that we analysed in more detail in this paper, with the superscripts giving their shorthand names (B2, B5, B6, B8, B10, S6, and S8). The other DNS were used for validation and for plotting the regime diagrams in the $(\theta,\mathrm{Re})$ plane.
	
	Finally, {we adopt a uniform} grid size $N_x \, \times \, N_y\,\times \, N_z$ for the entire domain $L_x\,\times \,L_y\,\times \,L_z$, {which has the advantage of helping maintain numerical stability near the immersed boundaries. The grid size was small enough to capture the Kolmogorov turbulent lengthscale, and $2-3$ times the Batchelor lengthscale in our most turbulent dataset B10 (discussed in more detail in \S\ref{sec:turb_fluxes}), ensuring adequate resolution of the kinetic and scalar energy spectra.}
	

	\section{Validation}\label{sec:valid}

\begin{figure}
		\centering		
		\includegraphics[width=.45\linewidth, trim=0mm 0mm 0mm 0mm, clip]{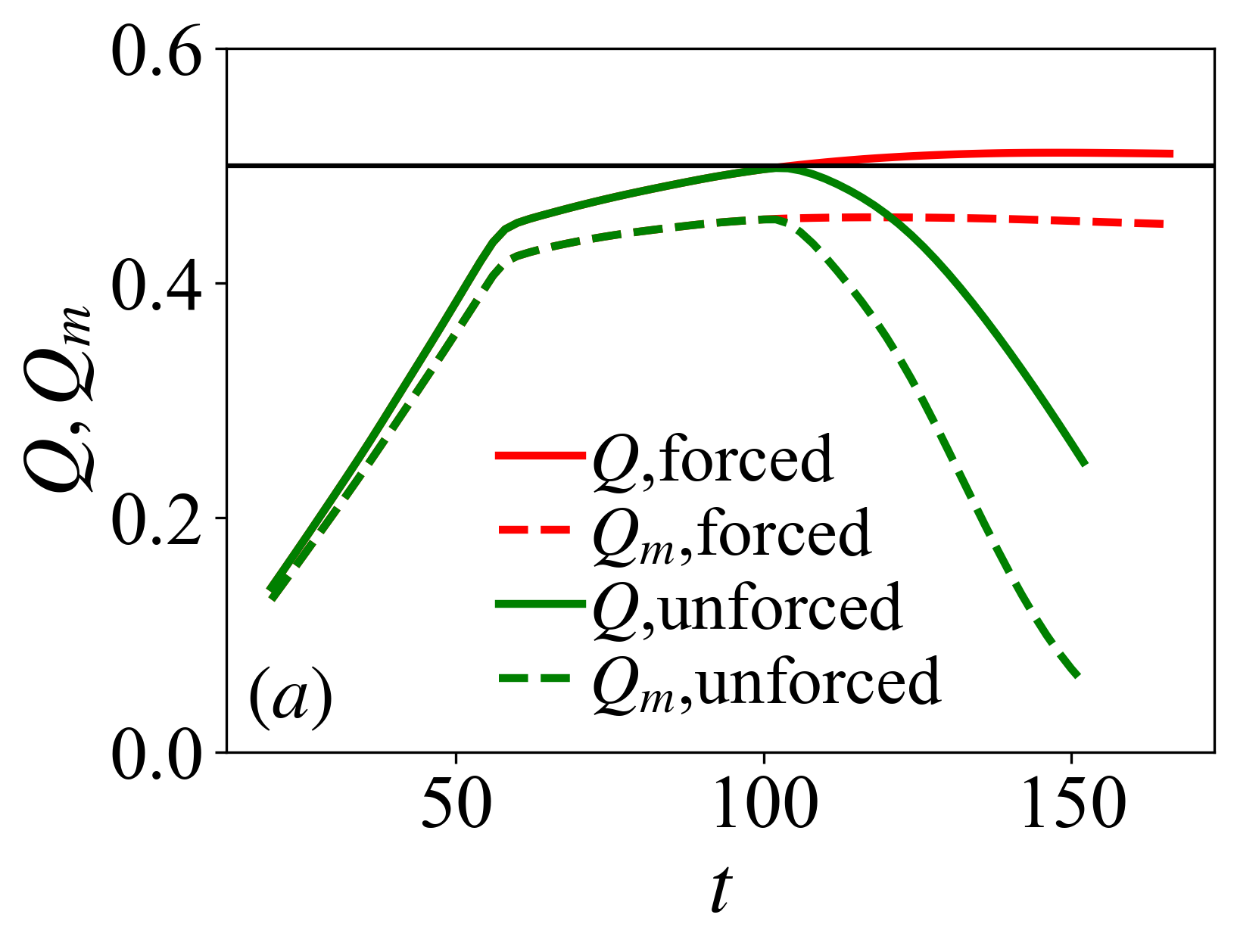}
		\includegraphics[width=.45\linewidth, trim=0mm 0mm 0mm 0mm, clip]{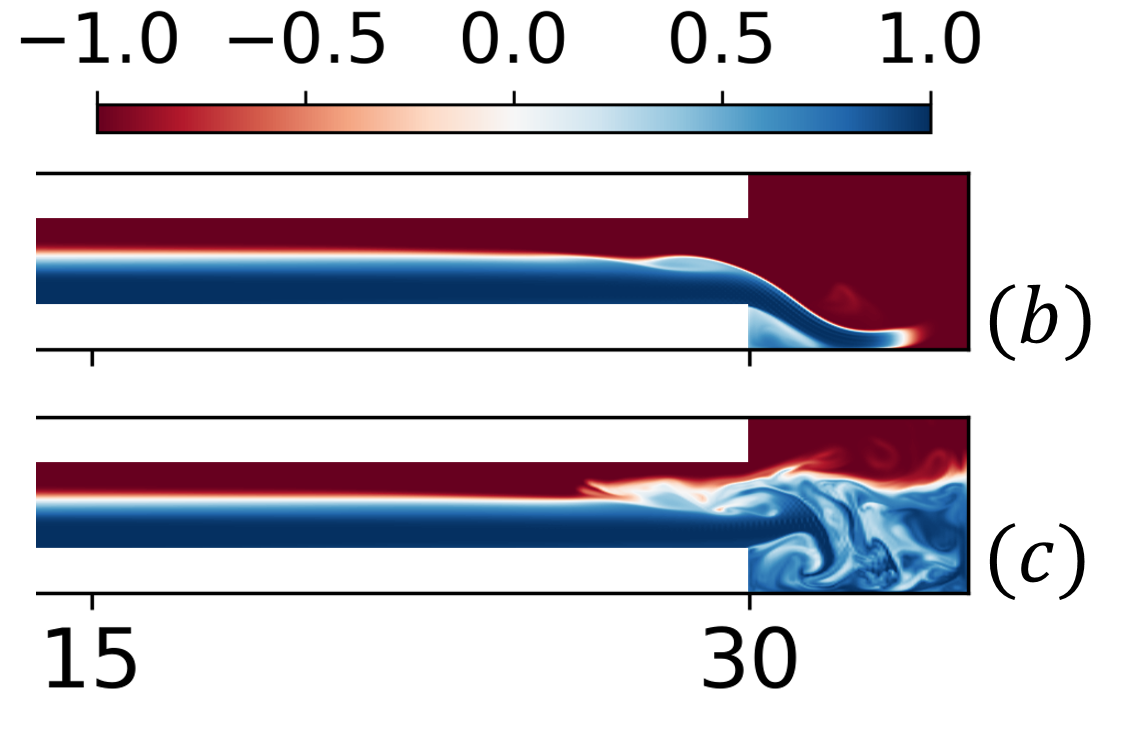}
		\caption{Demonstration of the effects of the forcing terms $F_u,F_\rho$ in finite-sized reservoirs (here in a B-reservoir, `BR') at $(\mathrm{Re},\theta)=(650,4^\circ)$. (a) Time series of the volume and mass flux. (b,c) Instantaneous mid-duct slices of $\rho(x,y=0,z,t=100)$: (b) forced and (c) unforced DNS, showing only the right-most quarter of the duct and the right reservoir.  }
		\label{fig:U_ts_com}
	\end{figure}

    In figure~\ref{fig:U_ts_com} we assess the ability of the forcing introduced in \eqref{eq:F_v}-\eqref{eq:F_ro} to sustain the exchange flow by comparing, for the B-reservoir, a standard DNS with forcing (`forced') and a DNS without forcing (`unforced') i.e. $F_u=F_\rho=0$. 
    
     Figure~\ref{fig:U_ts_com}(a)  shows the time series of the volume flux $Q$ and the mass flux $Q_m$ defined as
     \begin{eqnarray} \label{def-Q-Qm}
        Q(t)  &\equiv& \langle \vert u \vert\rangle_\mathcal{V}, \\
          Q_m(t)  &\equiv& \langle  \rho u \rangle_\mathcal{V},
     \end{eqnarray}
    where $\langle \cdot \rangle_\mathcal{V} \,\equiv\, (1/8A)\int_{-1}^1\int_{-1}^1 \int_{-A}^{A} \, \cdot \, dx \, dy \, dz$ denotes an average over the entire volume of the duct. Note that since $\vert \rho \vert \le 1$, by definition $Q_m\le Q$. A more diffuse interface and turbulent mixing can cause $Q_m$ to be significantly lower than $Q$. The values of $Q(t)$ (solid lines) and $Q_m(t)$ (dashed lines) in the forced (red) and unforced (green) DNS are identical in the initial stage of accelerating gravity current  ($0 < t \lesssim 60$). They remain equal until the exchange flow approaches a steady-state at $Q\approx 0.5$ and $Q_m\approx 0.45$ ($60\lesssim t \lesssim 100$). However, from $t\approx 100$, the unforced time series drops sharply, signalling that the flow slows down (see $Q(t)$) and becomes overall more mixed inside the duct (as $Q_m(t)$ decays faster than $Q(t)$). By contrast, the forced time series remains steady until the end  ($t\approx 160$) of the simulation. 
    
     Figures~\ref{fig:U_ts_com}\textit{(b,c)} show  $x-z$ slices of the density field on the rightmost quarter of the computational domain at $t=100$ for the forced DNS (panel b) and unforced  DNS (panel c). In the unforced DNS, the dense, right-flowing bottom layer (in blue) has filled over half of the B-reservoir. The large kinetic energy of this layer has led to mixing inside the reservoir. This dense fluid contaminates the exchange flow as it is entrained back into the duct by the left-flowing buoyant layer (in red). In the forced DNS, this does not happen; the outflowing layer is slowed down and its density is gradually converted to that of the inflowing fluid. This allows an infinitely-long quasi-steady exchange flow to be maintained inside the duct.
    
\begin{figure}
		\centering		
		\includegraphics[width=.99\linewidth, trim=0mm 0mm 0mm 0mm, clip]{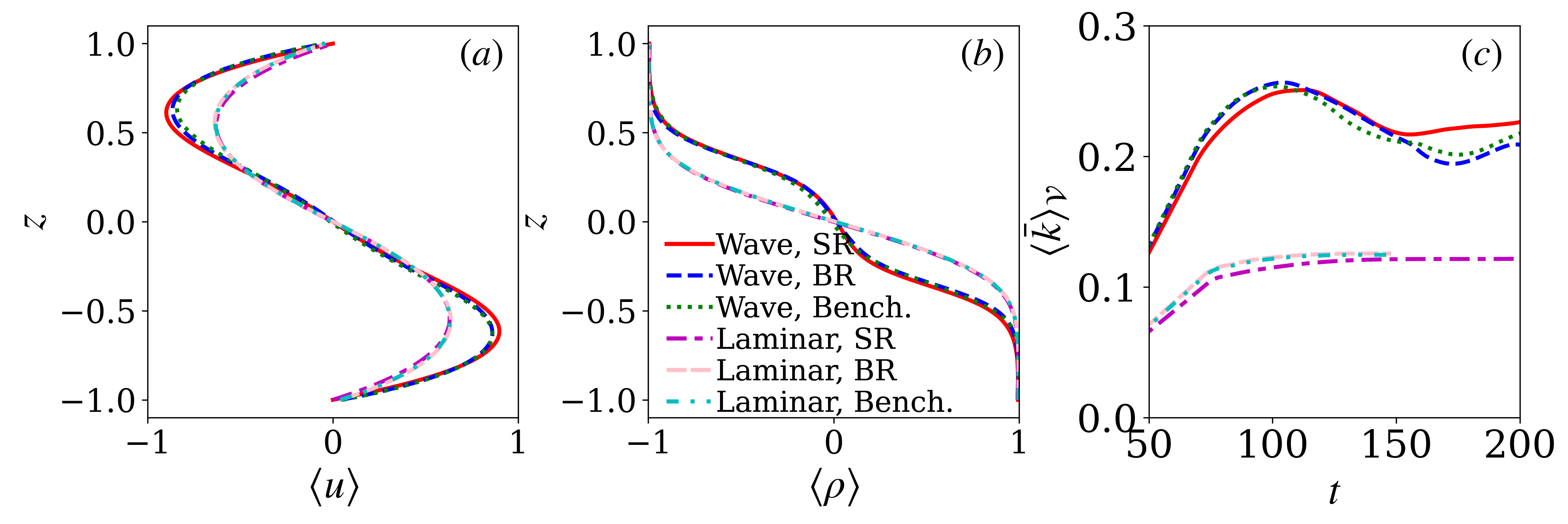}
		\caption{{Comparison of the effects of reservoir sizes on the: (a) streamwise velocity and (b) density profiles, and (c) kinetic energy time series in both the laminar and wave regimes.} } 
		\label{fig:stat_forc}
	\end{figure}

	In figure~\ref{fig:stat_forc} we compare the statistics of the established exchange flow in the Bench. case (very large reservoirs, unforced), and in progressively smaller, {but forced, reservoirs: BR and SR.} We compare two different flows: a laminar regime at $(\mathrm{Re},\theta)=(400,5)$ (red, blue, green curves) and a wave regime at $(\mathrm{Re},\theta)=(650,6)$  (purple, pink, and cyan curves). Panel (a) shows $\langle u \rangle(z)$ (where $\langle \cdot \rangle \equiv \langle \cdot \rangle_{x,y,t}$ is the average over the entire duct length, width, and time series), panel (b) shows $\langle \rho \rangle(z)$, and panel (c) shows the time series of the total kinetic energy $\langle \bar{k} \rangle_\mathcal{V}(t)$ (where $\bar{k}\equiv |\mathbf{u}|^2/2$). 
	
{Comparing the Bench., BR, and SR cases, we find excellent agreement between the vertical profiles and the time series of kinetic energy.  Minor temporal discrepancies in the wave regime after $t\gtrsim 150$ are probably caused by variations in the initial random noise, but have negligible influence on flow statistics and dynamics.  Overall, our forcing method faithfully models the effects of reservoirs as far as the flow inside the duct is concerned, even in very small reservoirs.} {We provide additional evidence that the key flow dynamics in the SID are largely independent of the reservoirs in Supplementary Material S1. We compare  spatio-temporal diagrams of the turbulent kinetic energy for $(\mathrm{Re},\theta)=(650,6^\circ)$ for the benchmark, AR, BR, as well as BR with a reservoir wider than the duct $L_y=4$ and show that details of wave motion and occasional turbulence over 200 advective time units do not vary more than they would under different initial noise conditions. }

Our typical computation at $\mathrm{Re}=650$ required $45\times 10^6$ points in the AR, but only $17\times 10^6$ in the BR (a reduction of 60\,\%). This explains why, in the following, we use the BR for more detailed analyses requiring longer time series of order $\approx 200$~ATU. We use the even more affordable SR more sparingly in this paper, since our main goal is to compare the SR results to the BR to investigate the ability of the SR to reproduce the key flow physics.

	\section{Comparison between DNS and experiments}\label{sec:transition}

	\subsection{Regimes: observations}
	
\begin{figure}
		\centering		
		\includegraphics[width=1.03\linewidth, trim=6mm 5mm 3mm 0mm, clip]{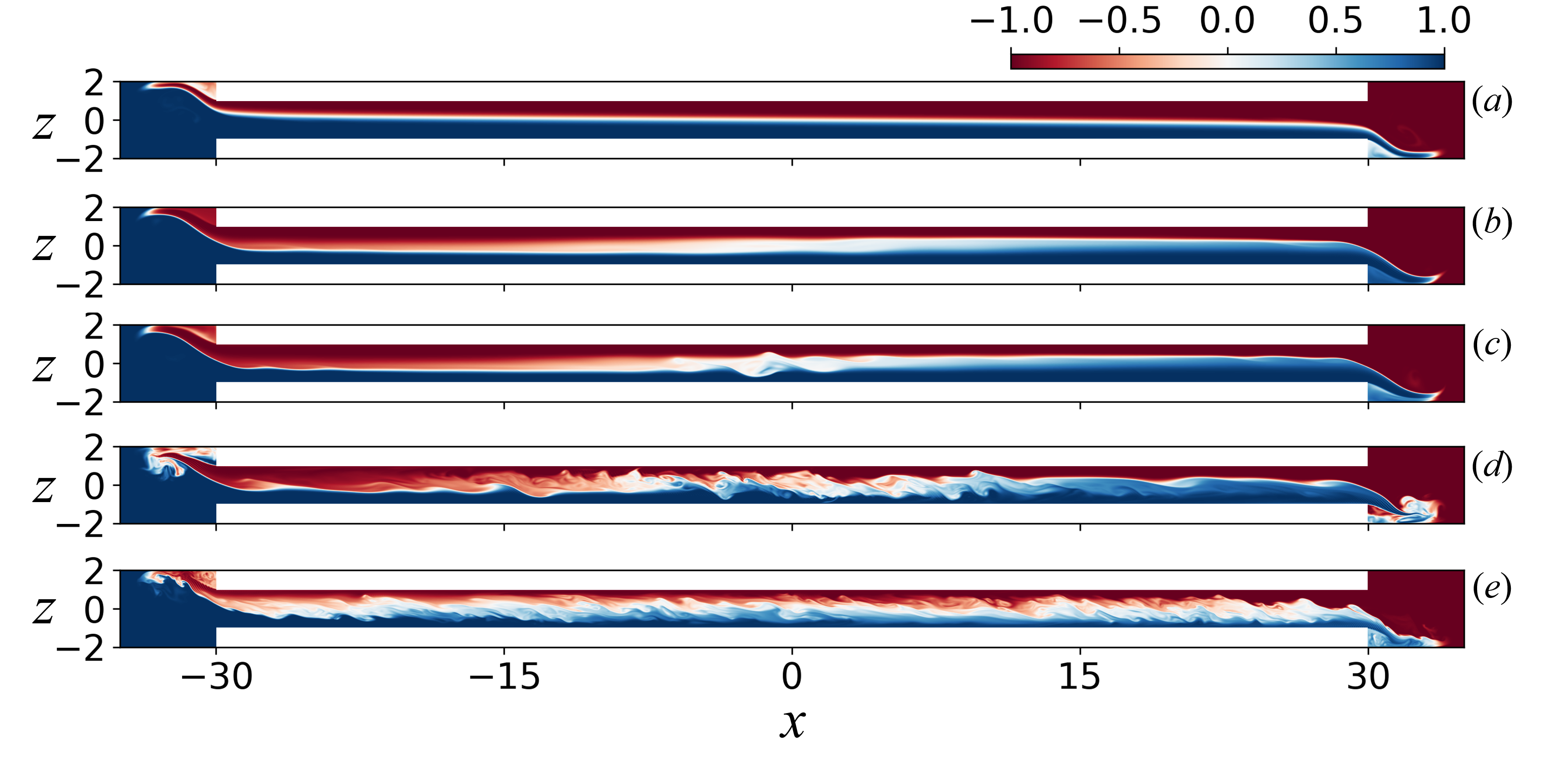}
		\\
		\includegraphics[width=.7\linewidth, trim=0mm 0mm 0mm 2mm, clip]{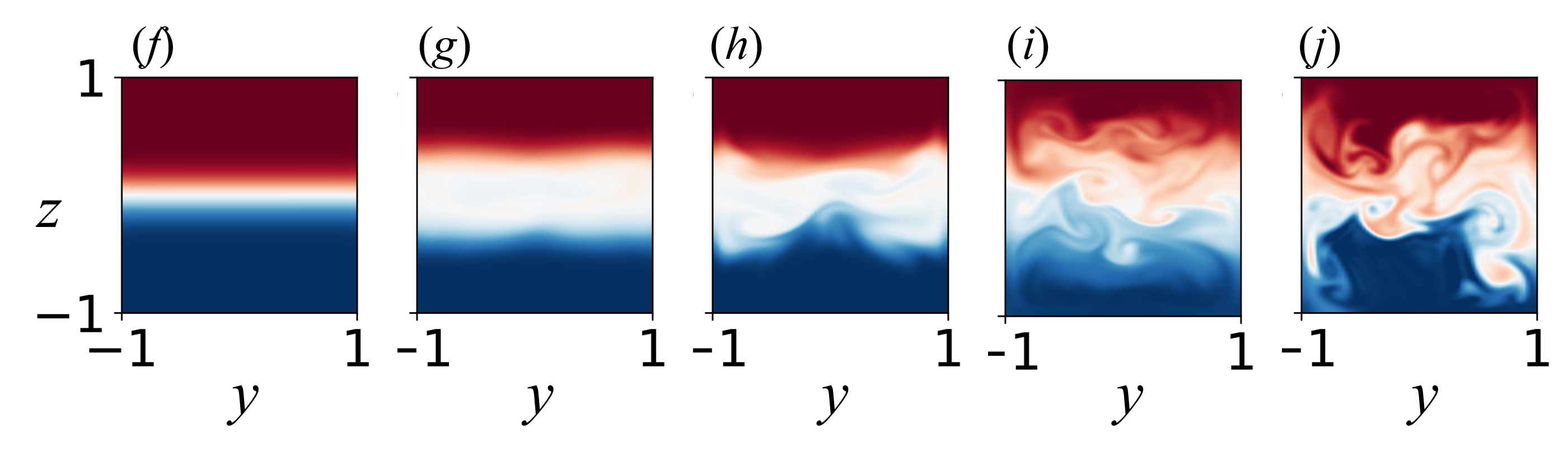}
		\caption{Snapshots of the density field in the mid-plane $y=0$ (top five panels) and in the duct cross-section (bottom five panels) in five representative flows: (a,f) laminar (B2), (b,g) stationary wave (B5), (c,h) travelling wave (B6), (d,i) intermittent turbulence (B8, active phase), and (e,j) fully-developed turbulence (B10). These cases are highlighted in bold font in \cref{tab:simul_overview}. 
		}
		\label{fig:ins_flow}
	\end{figure}

	Figure \ref{fig:ins_flow} shows snapshots of the DNS density field exemplifying the {quasi-}steady states of the different flow regimes.  All cases use the B-reservoir, with a duct aspect ratio of $A=30$, as highlighted in table~\ref{tab:simul_overview} and named B2, B5, B6, B8, and B10. The first four cases B2-B8 were at $\mathrm{Re}=650$, the last one B10 was at $\mathrm{Re}=1000$, and the numbers 2, 5, 6, 8, 10 indicate the respective value of $\theta$ in degrees. Slices through the density field at the middle $y=0$ plane (top five panels), and through a cross-sectional $y-z$ plane at $x=0$ in the duct are shown. The full temporal evolution of these five cases can be seen in our Supplementary Movies.
	
	We recover, in our DNS, the same four key flow regimes (laminar, wave, intermittently turbulent, and fully turbulent) identified in experimental studies of SID, in particular \cite{meyer2014stratified,lefauve2018waves,lefauve2019regime,lefauve2020buoyancy}, which is a first key result of this paper. Moreover, the DNS allow us, for the first time, to observe three-dimensional instantaneous snapshots along the entire domain, including the duct and the in- and out-flow in the reservoirs, which were not accessible to experiments. We describe each regime in turn.
	
	\subsubsection{Laminar regime}
	First, in B2 (figure~\ref{fig:ins_flow}(a,f)) we observe a simple laminar (L) flow, which is largely parallel and steady without any observable waves or turbulent fluctuations. Molecular diffusion creates a relatively thin interface of intermediate density (in white). This density interface slopes at an angle since the two counter-flowing layers (in blue or red) get thinner as they accelerate along the duct. {This convective acceleration  $u\partial_x u$ in each layer is caused by the pressure gradient $-\partial_x p$, gravity $\textrm{Ri}\,  \sin\theta \,\rho$ and opposed by  the viscous term $\mathrm{Re}^{-1}\boldsymbol{\nabla}^2 u$. }
	
	\subsubsection{Wave regime(s)}
	Second, in B5 and B6 (figure~\ref{fig:ins_flow}(b,c,g,h)) we observe a wave (W) flow, including large-scale waves with streamwise wavelength of  order $O(1) - O(A)$ (i.e. $O(1) - O(30)$). These waves are triggered by disturbances within the duct; information (waves)  does not travel from the reservoirs into the duct. 
	In this regime, small-scale, weakly-turbulent structures of limited spatial extent may occasionally be generated by the breakdown of large-scale waves but they always dissipate rapidly. We note that most of the previous SID experiments were done with salt stratification ($Pr\approx700$), in which case the much-thinner density interface supports Holmboe waves; hence these studies called this wave regime the `Holmboe' regime. However, with temperature stratification  ($\textrm{Pr}\approx7$, as simulated here) \citet[][\S~3.4.2]{lefauve2018waves} and \citet{lefauve2020buoyancy} highlighted that Holmboe waves were never found on the thicker interface. At $\textrm{Pr}\approx 7$ they found the same wave regime  observed here, with interfacial gravity waves on the edges of a thicker density interface. 
	
	We find that W flows can feature  waves that are either stationary (B5; figure~\ref{fig:ins_flow}(b,g)) or travelling (B6; figure~\ref{fig:ins_flow}(c,h)) in the streamwise direction $x$. Increasing $\theta$ tends to first decrease the slope of the interface (relative to the duct) and accelerate the flow until the interface is parallel to the axis of the duct (reducing $u\partial_x u$ and $-\partial_x p$), at which point the gravitational term $\mathrm{Ri} \sin \theta \rho$ can no longer be balanced by laminar diffusion alone. This appears to coincide with the creation of a third, partially mixed layer (in light red, white, and light blue) that is neutrally buoyant and thus reduces the gravitational forcing $\mathrm{Ri} \sin \theta \rho$. This third layer, often located near the centre of the duct (i.e. around $|x|\approx 0$) rather than near the ends ($|x|\approx A$) supports both stationary and travelling interfacial waves. We note that, in contrast to L flow, in these W flows the in-flowing layers (before reaching the `wavy' area in the centre of the duct) are thinner than the out-flowing layers (after going through the `wavy' area), which is somewhat reminiscent of an internal hydraulic jump. 
	
	Travelling waves (figure~\ref{fig:ins_flow}(c)) tend to travel along the two density interfaces (between the top and middle layers, and between the middle and bottom layers) in a specific fashion. Left-going waves are most often found on the right quarter of the duct ($x \gtrsim A/2$), travelling towards the centre. Vice versa, right-going waves are most often found on the left quarter of the duct ($x \lesssim -A/2$), travelling towards the centre. Once they reach the central region  ($|x| \lesssim A/2$), both types of waves usually end up decaying. This  observation suggests that the flow may be `supercritical' outside of the central region, i.e. that information transported by interfacial waves can only propagate in one direction (towards the centre but not toward the ends).

	\subsubsection{Intermittently turbulent regime}
	Third, in B8 (shown in figure~\ref{fig:ins_flow}(d,i)) we observe an intermittently turbulent (I) flow, which becomes more chaotic and in which patterns of individual waves become indistinguishable. Small-scale  turbulent structures (of typical non-dimensional scale $\ll 1$) are generated, often by a breakdown of waves akin to the `bursting' events of turbulent boundary layers~\citep{robinson1991coherent,jimenez2001low,Zhu_XiJNNFM2020}. This interfacial turbulence, which persists for much longer times than in the W regime, enhances interfacial mixing and creates a third partially mixed layer (shown in white) over an increasingly long streamwise extent (as compared with the W regime). The interfacial turbulence sometimes extends along the full length of the duct. The combination of the decreasing magnitude of the gravitational forcing $\mathrm{Ri} \sin\theta |\rho|$ by the increasingly mixed layer and the increasing smaller-scale viscous dissipation are presumably the key ingredients that keep the flow steady as  $\theta$ is increased from $2^\circ,5^\circ,6^\circ$ to $8^\circ$ in B2, B5, B6, B8 (at constant $\mathrm{Re}=650$).
	
	The defining characteristic of the I regime is that the turbulence identified by small-scale structures is temporally intermittent; turbulence occasionally decays and the flow `relaminarises' before transitioning to turbulence again; these cycles will be described  in \S\ref{sec:flowDyn}. The time scales associated with the transition to turbulence and its decay, and the advection of perturbations along the length of the duct, occasionally make this turbulence also spatially intermittent in $x$.
	
	\subsubsection{Fully turbulent regime}
	 Fourth, in B10 (shown in figure~\ref{fig:ins_flow}(e,j))  we observe a fully turbulent regime (T) in which turbulence is sustained in time and is more vigorous than in the I regime. Although the intensity of the turbulence can fluctuate in time, the flow in this regime never fully relaminarises. The central partially-mixed layer typically covers the entire length of the duct and at least a third of the height of the duct.

	\subsection{Regime diagrams}

\begin{figure}
		\mbox{\emph{(a)} \ \ A-reservoir}  \hspace{5cm}	\mbox{\emph{(b)} \ \ B-reservoir} \\ 
			\centering
		\includegraphics[width=\textwidth]{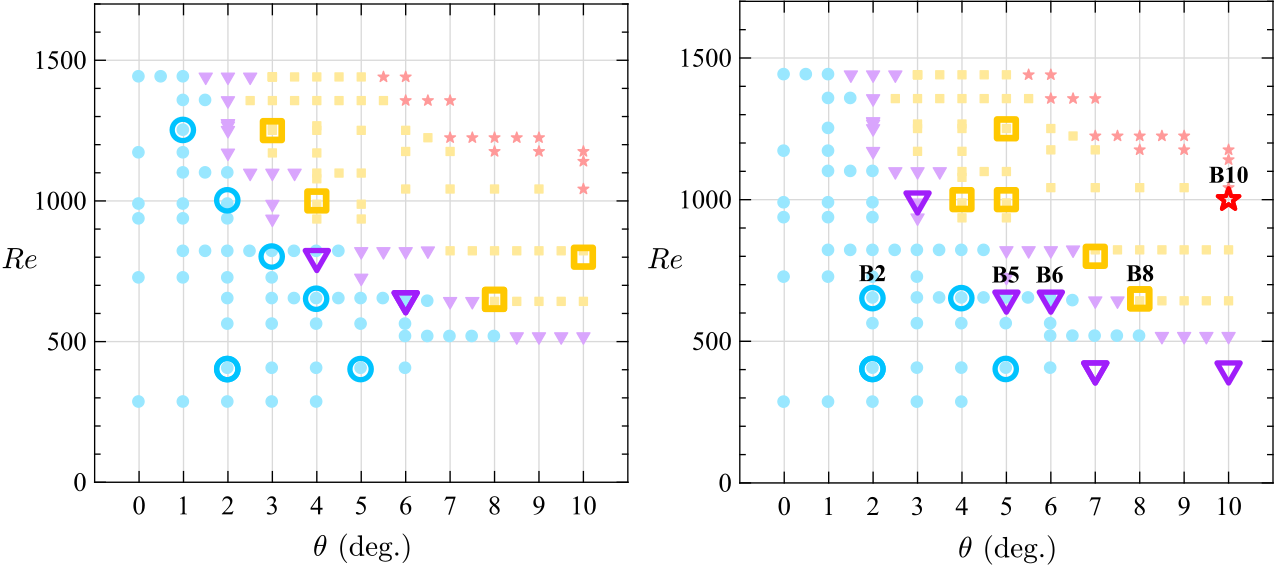}
		\includegraphics[width=0.5\textwidth]{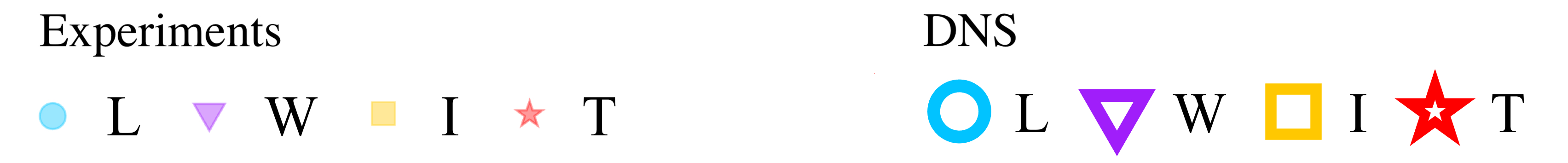}
		\caption{Regime transitions in $\theta-\mathrm{Re}$ parameter space.  The large symbols are our DNS data in the (a) A- and (b) B-reservoir, and the small markers are the temperature-stratified experimental data of \citet{lefauve2020buoyancy} (see their figure 4\textit{e}) with matched non-dimensional parameters.}
		\label{fig:regime}
	\end{figure}

	In figure~\ref{fig:regime} we map the flow regimes described above in 12 DNS with the AR (panel a) and in 15 DNS with the BR (panel b) for a range of $\theta$ and $\mathrm{Re}$. 
	
	These 27 DNS data points, shown as large symbols, are compared with the 148 experimental data points of \cite{lefauve2020buoyancy} taken from their figure 4(\emph{e}) and displayed here as smaller, fainter symbols using the same colour coding for the different flow regimes. The experimental data points were obtained using the same aspect ratios ($A=30, B=1$) and with temperature stratification ($\mathrm{Pr}\approx 7$). The regimes were identified by shadowgraph visualization, often over a small streamwise extent of the duct (their movies can be downloaded from \cite{lefauve_research_2020}).  \Cref{fig:regime} therefore represents the first direct comparison of DNS results with experimental results in SID, with all non-dimensional control parameters matched. 
		
	First, we find a general agreement between DNS and experiments in the location of flow regimes in the $(\theta,\mathrm{Re})$ plane, as evidenced by the fact that most large symbols (DNS) are of the same type as the smaller symbols (experiments). This is a second key result of this paper, because it confirms that our DNS, with small computationally-efficient reservoirs, can reproduce the key physics of SID, encapsulated in the flow regimes. 
	
	The minor exception to this agreement is found near the L/W transition, where some of our DNS found the W regime whereas the experiments found the L regime. This may be a genuine difference, but we suspect that this may be due to the fact that the weak stationary waves found near the L transition may have been missed in the experiments. This is because the experimental shadowgraphs were visualised over a limited extent of the duct and because low-amplitude waves in low-$\mathrm{Re}$ temperature-stratified flows produce very small changes in refractive index and thus weak shadowgraph signals.
	
	Second, the AR and BR yield consistent results (panels a and b), confirming that the smallest `true' reservoir (excluding the SR for now) is indeed sufficient to reproduce the experiments. These results offer strong further support to the preliminary validation of our suite of DNS in \S\ref{sec:valid}.

	\subsection{Shadowgraphs}\label{sec:shadowgraph}

We now turn to a side-by-side comparison of shadowgraph visualisations of the flow in DNS and experiments  within a particular flow regime.

Experimental shadowgraph movies are obtained by the projection onto a semi-transparent screen of initially parallel light rays that have travelled through the duct along the spanwise $y$-direction. Any variations in the curvature (normal to the rays) of the density field $\rho$ (and hence refractive index field $n$) causes  the rays to focus or defocus, varying the intensity that reaches the screen \citep{weyl1954analysis}. In the limit of weak variations, the intensity of the image formed is (see e.g. \citealp[\S~2.1]{lefauve2018waves})
%
%
\begin{equation}\label{eq:sg}
    I(x,z,t) = \beta I_0(x,z) \int_{-B}^{B} (\partial_{xx}+\partial_{zz})\rho(x,y,z,t) \,dy.
\end{equation}
Here $\beta$ depends on $(\rho_0/n_0)\partial n/\partial\rho$ and the experimental geometry,  while $I_0$ is the (approximately) uniform background intensity of the illumination.
%
%
This field is thus particularly suited to detect density interfaces, and is a simple and efficient proxy to compare the structure of interfacial density waves and small-scale turbulence in DNS and experiments. 

\begin{figure}
	    \centering
	    \hspace{-0.8cm}
		\includegraphics[trim=4mm 0 0 0 ,clip,width=1.04\textwidth]{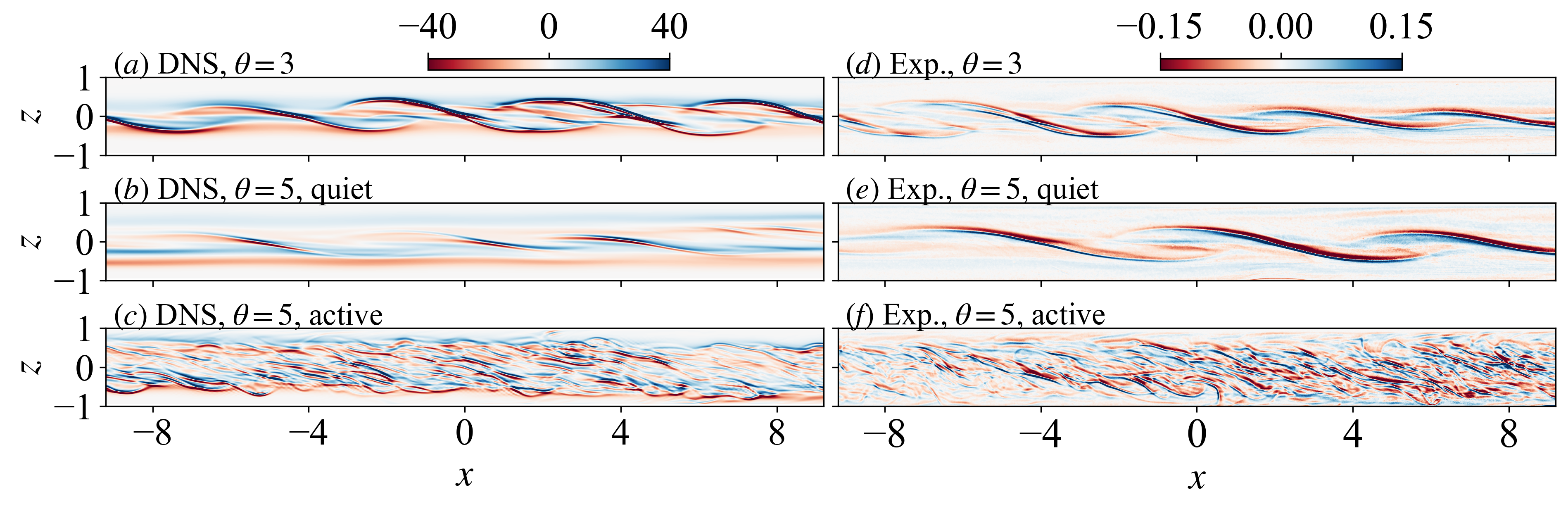}
		\caption{Snapshots of shadowgraph comparing the normalised intensity $I(x,z)$ in DNS (left column)   to the matching experiments (right column) in two cases W3 (top row) and W5 (bottom two rows). Magnitudes (colour bar limits) are naturally different  due to the unknown experimental $\beta$ factor in \eqref{eq:sg}. The times at which these snapshots were taken are shown by the vertical lines in the spatio-temporal diagrams of figure~\ref{fig:SG_zt}.}
		\label{fig:SG}
	\end{figure}

\Cref{fig:SG} compares false colour instantaneous snapshots of $I(x,z)$ in the I regime over a central portion of the duct $|x|<9$. The DNS shadowgraphs reconstructed from the calculated density fields (assuming $\beta I_0 =1$) are shown in the left column (W3 and W5, see \cref{tab:simul_overview}), and the matching experimental shadowgraph images of $I/I_0$ are shown  in the right column, all at $\textrm{Re}=650$ and $\textrm{Pr}=7$. 
We show a single snapshot at $\theta=3^\circ$, at the boundary between the W and I regime (panel a) and two snapshots at $\theta=5^\circ$, well into the I regime, where the flow is in a quiet laminar phase (panel b) and in an active turbulent phase (panel c). The full temporal evolution of these four shadowgraphs can be found in our Supplementary Movies.

Note that these shadowgraphs were obtained in a new experimental apparatus having a wide duct $B=2$, with a regular straight rectangular section of length $A=40$, and trumpet-shaped expansions at either end (over an additional 10~\% of its length) for a smoother connection to the reservoirs. While we did not model the trumpet ends in our DNS, we used $B=2$ and the total length $A=44$ to reproduce the geometry as faithfully as possible. Trumpet ends were first used in \cite{meyer2014stratified} who reported no visible impact in shadowgraphs when compared to straight ends.   With  the parameters $A$ and $B$ increased with respect to cases B2-10, the I regime is found at smaller $\theta$ values than would be expected from figure~\ref{fig:regime}  (see \cite{lefauve2020buoyancy}).

We find good agreement in the structure of interfacial waves, somewhat reminiscent of Kelvin-Helmholtz billows, in DNS and experiments (compare panels a,b to d,e). These waves have higher amplitude than the stationary waves previously found in B5 (see figure~\ref{fig:ins_flow}(b)) because the flow is more energetic and prone to the growth of stratified shear instabilities at $B=2$ than at $B=1$, due to a weaker influence of the no-slip side walls (see \citealp[\S~IIIc]{ducimetiere2021effects}). These waves tend to break into  weak and short-lived turbulence at $\theta=3^\circ$ (placing it borderline in the I regime), and into stronger and longer-lived turbulence at $\theta=5^\circ$ (placing it well into the I regime).
We also find good agreement in the overall appearance of small-scale turbulence in the `active' phase (compare panels c to f). 
{Active turbulence in the experiment extends slightly closer to the top and bottom boundaries than in the DNS. This may be a result of various factors including the non-zero thermal conductivity of the experimental duct walls, spurious reflections of light, and excessive cropping of near-wall regions caused by the difficulty in locating the wall in the shadowgraph images.}

\begin{figure}
	    \centering
		\includegraphics[width=0.68\textwidth]{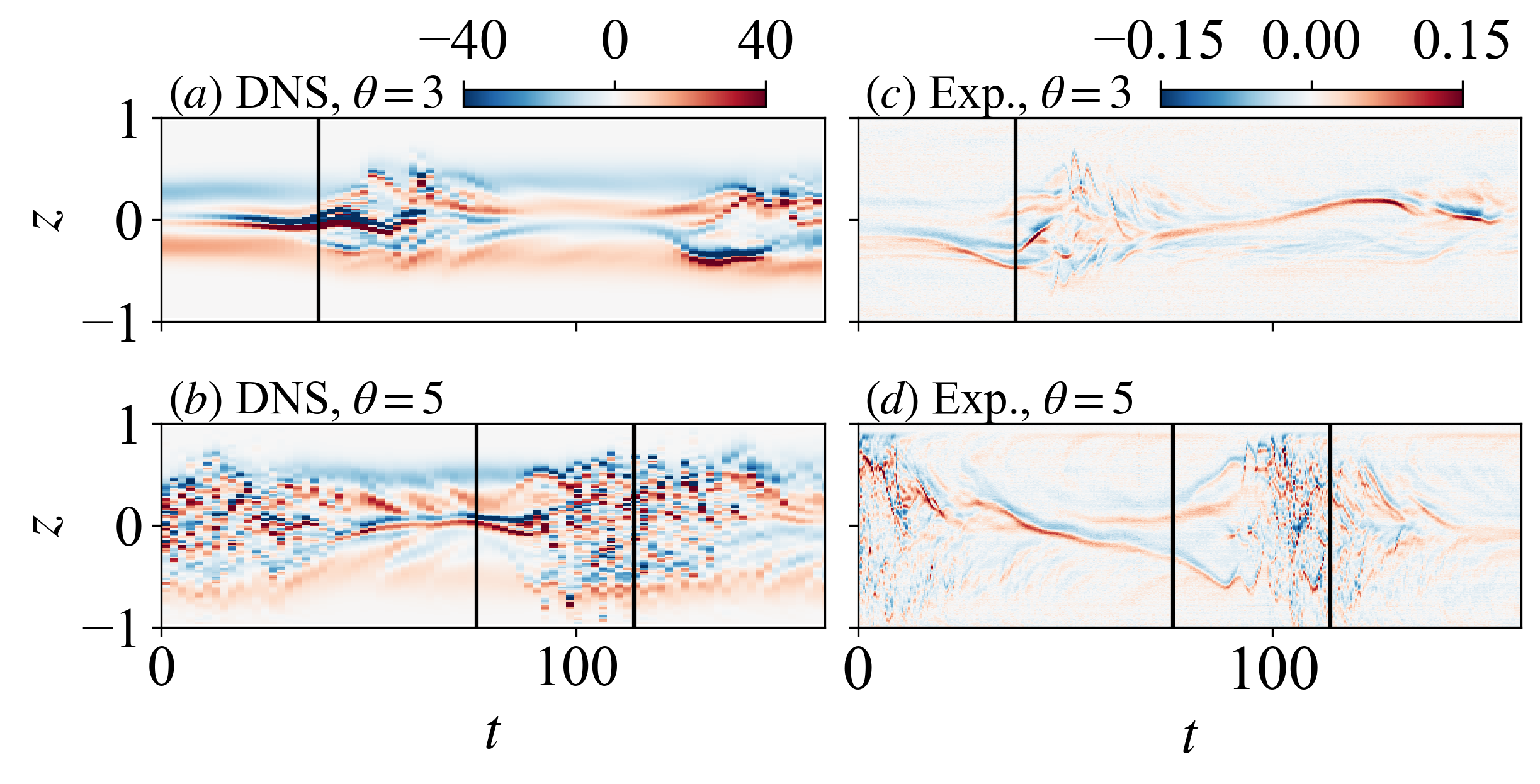}
		\caption{Spatio-temporal diagrams of shadowgraph in  W3 (top row) and W5 (bottom row) comparing DNS (left column)  to experiments (right column). The vertical black solid lines indicate the time of the snapshots in figure~\ref{fig:SG}.}
		\label{fig:SG_zt}
	\end{figure}

\Cref{fig:SG_zt} illustrates these temporal dynamics  with the corresponding $z-t$ spatio-temporal diagrams in DNS (left column) and in experiments (right column). We find again good agreement, both in the vertical growth and decay of the waves, and in the alternation and approximate period of the quiet and active phases. 

These shadowgraphs show that our DNS faithfully reproduce not only the qualitative flow regimes and their distribution in $\theta-\textrm{Re}$ space, but also details of their spatial structures and temporal dynamics, which is a third key result of this paper.

	\section{Added value of DNS}\label{sec:addedvalue}

 In this section we examine  quantitative DNS diagnostics  which, because they are difficult or impossible to obtain in experiments, add value to the experimental study of SID.
    
	
	\subsection{Vertical profiles and gradient Richardson number}

\begin{figure}
		\centering		
	\hspace{-1cm}	\includegraphics[width=1.02\linewidth, trim=0mm 0mm 0mm 0mm, clip]{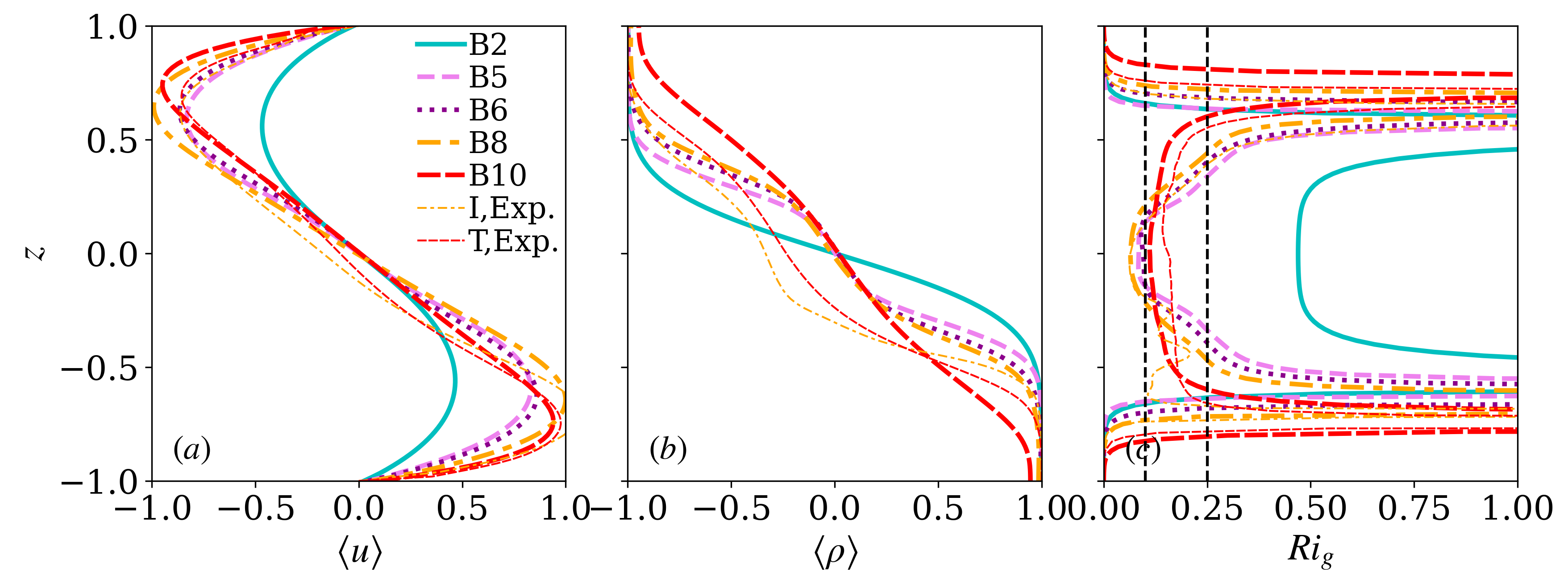}
		\caption{Vertical profiles of the mean (a) streamwise velocity $\langle u \rangle$, (b) density $\langle \rho \rangle$ and (c) gradient Richardson number $\mathrm{Ri}_g$, in the five flows of figure~\ref{fig:ins_flow}. We also {include} the I and T experimental profiles in \cite{lefauve2019regime} at $\mathrm{Pr}\approx 700$. The vertical dashed lines in (c) denote $Ri_g=$ 0.1 and 0.25.  }\label{fig:Stat_case}
	\end{figure}

	\Cref{fig:Stat_case} shows, for the five  flow regimes previously shown in figure~\ref{fig:ins_flow}, the  $x,y,t$-averaged velocity $\langle u \rangle(z)$, density $\langle \rho \rangle(z)$, and the gradient Richardson number $\mathrm{Ri}_g$ based on the gradients of these mean flows
	\begin{equation}
		\mathrm{Ri}_g(z)\equiv-\mathrm{Ri}\,\frac{\partial_z \langle \rho \rangle}{(\partial_z \langle u \rangle)^2}.
	\end{equation}
	
Such simultaneous velocity and density diagnostics are available in salt-stratified experiments (at $\mathrm{Pr}\approx 700$), and  we superimpose on figure~\ref{fig:Stat_case} the mean profiles in the I and T regimes {from} \cite{lefauve2019regime} (their figure 4\textit{f,l}). However, these diagnostics cannot be accurately obtained in temperature-stratified experiments (at $\mathrm{Pr}\approx 7$) to match our DNS for two main reasons. First, the velocity field measurements rely on particle image velocimetry (PIV) in a refractive index matched fluid, which is impossible without the introduction of another stratifying agent (necessarily having a much smaller diffusivity than temperature). Second, because the density field measurements rely on laser-induced fluorescence (LIF) with a dye having a much smaller diffusivity than temperature, and therefore `tagging' it poorly (temperature-sensitive fluorescent dyes exist but such measurements are more difficult and less accurate).


	In figure~\ref{fig:Stat_case}(a), the velocity in the L regime (B2) {adopts an approximately} sinusoidal profile with a low amplitude ($\max |u| \approx 0.3$), whereas in the SW, TW, I and T regime{s} (B5, B6, B8, B10, respectively), the mean velocity varies nearly linearly with height for $|z|\lesssim 0.5$, and $\max |u| \approx 1$. The vertical locations of the peaks in velocity, initially around $z \approx \pm 0.5$ in the L regime,  shift slightly towards the top and bottom walls $z \approx \pm 0.7$ in the I and T regimes. These observations 
	{agree qualitatively with}  
	 the experimental profiles of \cite{lefauve2019regime} in the four regimes (see their figures 3\textit{(f,l)} and 4\textit{(f,l)}).
	{However, exact agreement should not be expected as their $\theta$, $\mathrm{Re}$ and $\mathrm{Pr}$ values differ from ours.  }
	
	In figure~\ref{fig:Stat_case}(b), the density profile resembles an error function in the L regime (B2) whereas it has a partially mixed layer in the W and I regimes (B5-8 and I,Exp.) identifiable by a central region of reduced gradient (a layer) flanked by two regions of enhanced gradient (two interfaces). These three profiles are almost identical, with the nuance that the intermediate layer becomes slightly thicker from B5 to B6 to B8, as expected.  Finally, in the T regime (B10 and T, Exp.), the middle layer becomes noticeably thicker, and the two interfaces flanking it become less sharp, leading to a profile approaching a uniform stratification. 
	
	
   In figure~\ref{fig:Stat_case}(c), the L flow has $\textrm{Ri}_g\approx 0.5$ throughout the central quarter of the height of the duct, flanked by steeply increasing values. The W and I flows have $\textrm{Ri}_g\approx 0.1$ near $z=0$. The T flow has a broader minimum with $\textrm{Ri}_g\approx 0.1-0.15$. The $\textrm{Ri}_g$ profiles in the I and T flows are {qualitatively} consistent between DNS and experiments, showing that despite the difference {in parameters}, some key dynamical features of turbulence are not sensitive to the fluid properties. 
    
    Note that $\mathrm{Ri}_g<0.25$ at $z=0$ in the W, I, and T flows, but not in the L flow. Therefore, the mean profiles in the non-laminar flows reassuringly satisfy the Miles-Howard criterion necessary for the development of instabilities in a steady, inviscid, Boussinesq, parallel stably-stratified shear flow. Moreover, in the T regime $\mathrm{Ri}_g(z) \approx\mathrm{Ri}_e \approx 0.1-0.15$ (i.e.\ robustly below the Miles-Howard criterion of 0.25). This agrees with the experimental conclusions of \cite{lefauve_experimental1_2022} (see their figure 5), drawn from a wider data set of 16 flows with increasing levels of turbulence. 
    The reasons for this particular equilibrium, originally suggested by \cite{turner_buoyancy_1973} and much observed since in numerical, experimental and observational data, are still debated. {Other authors~\citep{thorpe_2010,smyth2013marginal,salehipour_self_2018}} have called it `self-organised criticality' or `marginal stability', and found values ranging from $\mathrm{Ri}_e\approx 0.07$ to $0.25$ in various stratified shear flows that differ (sometimes significantly) from SID \citep{lefauve_experimental1_2022}.

	\subsection{Kinetic energy}\label{sec:flowDyn}

   We now study the spatio-temporal dynamics of kinetic energy along the entire length of the duct in our DNS. Similar experimental diagnostics are not yet available since the resolution of video cameras and geometry of the laser sheet limit us to shorter windows spanning only a limited part of the duct length. 

\begin{figure}
		\centering		
		\includegraphics[width=.75\linewidth, trim=0mm 0mm 0mm 0mm, clip]{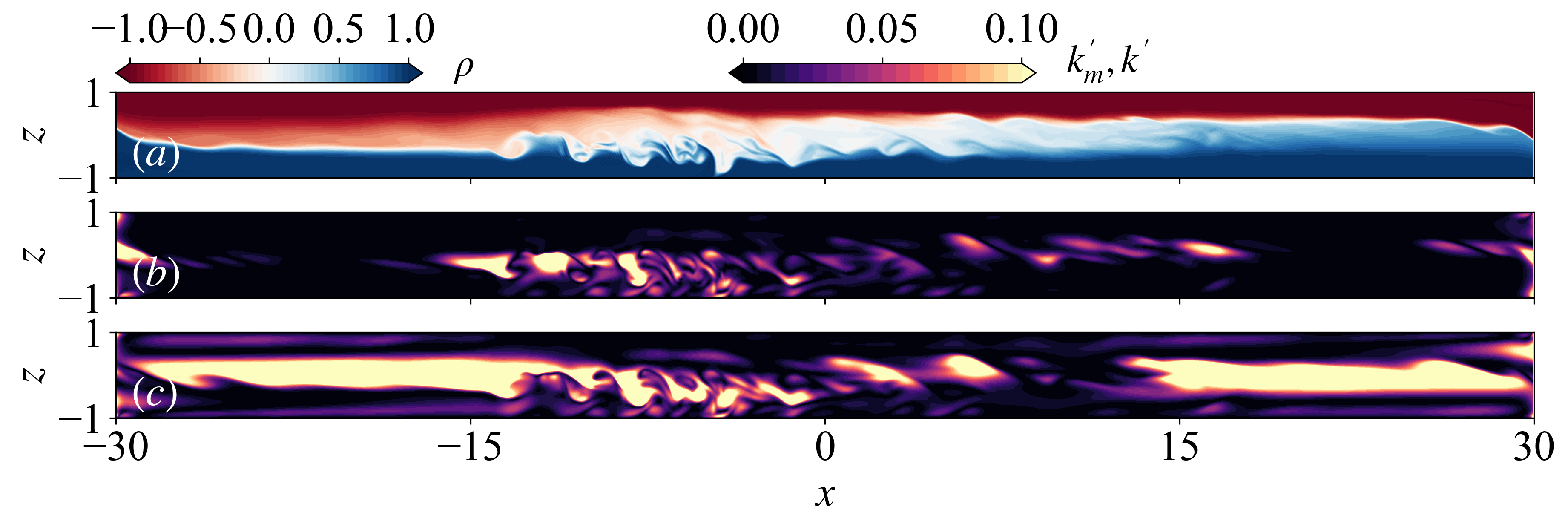}
		\includegraphics[width=.24\linewidth, trim=0mm 0mm 0mm 0mm, clip]{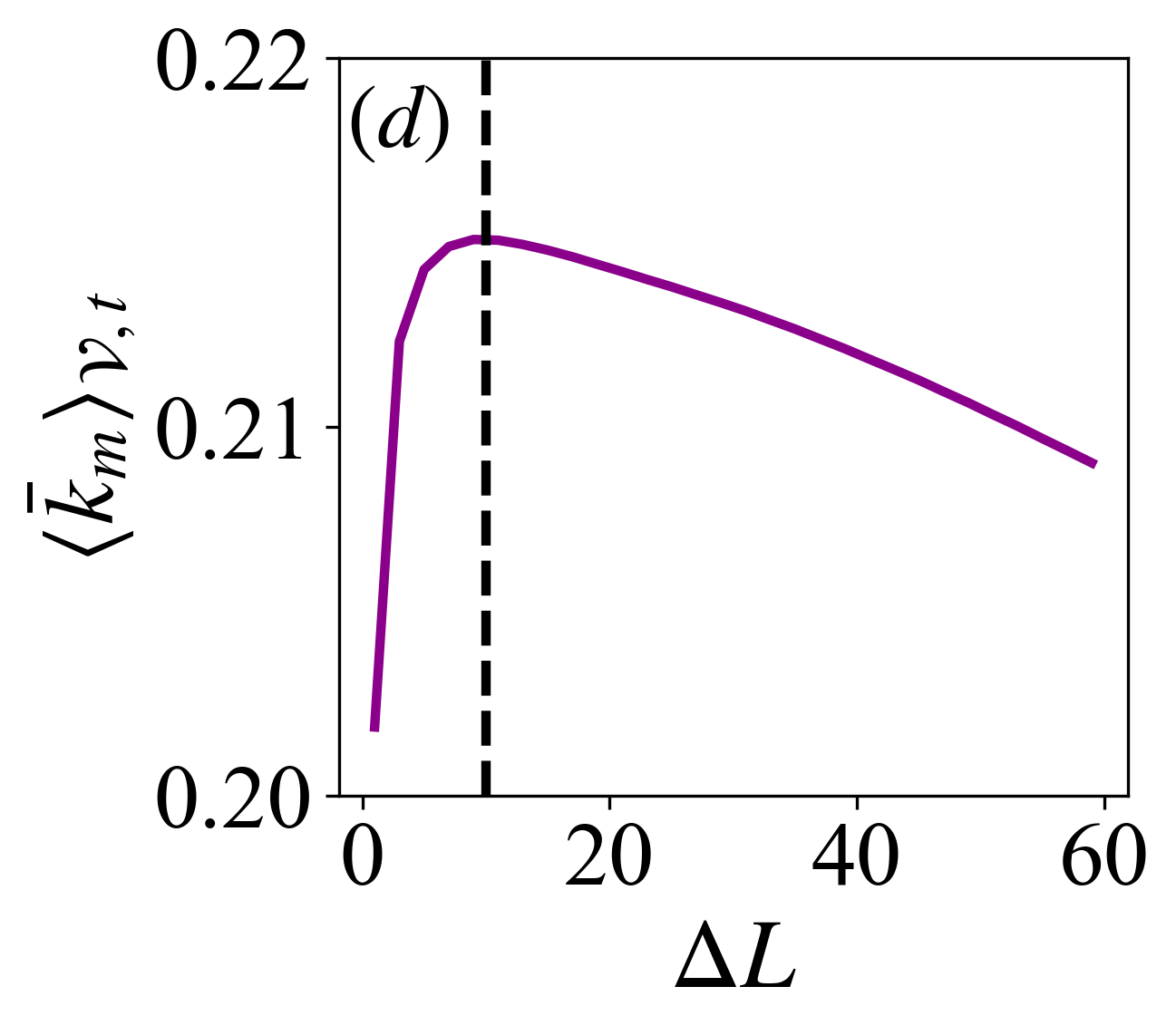}	
		\caption{Snapshots of DNS B6 at $t=160$ showing the (a) density, (b) moving-averaged TKE $k^\prime_m$ defined in \eqref{eq:mv:tke}, and (c) `naive' TKE based on $\mathbf{u}'=\mathbf{u}-\langle \mathbf{u}\rangle_{x,t}$; (d) bulk MKE $\langle \bar{k}_m \rangle_{\mathcal{V},t}$ as a function of the stencil length $\Delta L$. The dashed line corresponds to our choice in the remainder of the paper $\Delta L=10$.  }
		\label{fig:tke_window}
	\end{figure}

	We start by decomposing the velocity into mean and turbulent (fluctuating) components. \cite{lefauve_experimental2_2022} defined the mean as the $x,t$ average. However, as our DNS data are available {along} the entire length of the duct, the flow  (especially $u$) becomes noticeably inhomogeneous in $x$ (see figure~\ref{fig:ins_flow}(a-c)). A simple $x$ average would therefore make the `turbulent' component artificially large by incorporating a significant non-parallel -- but laminar -- component. To resolve this, we define the mean and fluctuations using a moving-average,
	\begin{eqnarray}
		\mathbf{\bar{u}}_m\left(x,y,z,t\right)&\equiv&\frac{1}{\Delta L}\int_{-\Delta L/2}^{\Delta L/2}{\mathbf{u}\left(x-s,y,z,t\right)ds},
		\label{eq:mv:um}
		\\
		\mathbf{u}^\prime_m(x,y,z,t)&\equiv&\mathbf{u}-\mathbf{\bar{u}_m},
		\label{eq:mv:uf}
	\end{eqnarray}
	and the respective `moving' mean kinetic energy (MKE) and turbulent kinetic energy (TKE) are
		\begin{eqnarray}
		\bar{k}_{m}\equiv\frac{1}{2}\mathbf{\bar{u}}_m \cdot \mathbf{\bar{u}}_m, \label{eq:mv:mke}
		\\
		k^\prime_{m}\equiv\frac{1}{2}\mathbf{u}^\prime_m \cdot \mathbf{u}^\prime_m.
		\label{eq:mv:tke}
	\end{eqnarray}
	The length of the averaging stencil $\Delta L=10$ was chosen to maximise the time- and duct-volume-averaged MKE $\langle \bar{k}_{m}\rangle_{\mathcal{V},t}$, as shown in figure~\ref{fig:tke_window}(d).

	In figure~\ref{fig:tke_window}(a-c) we demonstrate the use of this moving average with a snapshot in the travelling wave regime (B6, as in   figure~\ref{fig:ins_flow}(c)). The underlying turbulence `hotspots' visualised by the density field (panel a) are faithfully captured by our moving-averaged TKE $k^\prime_{m}$ (panel b),  whereas they are greatly overestimated by the `naive' TKE based on the $x,t$-averaged velocity (panel c), which is equivalent to setting $\Delta L=2A=60$ and $x=0$ in \eqref{eq:mv:um}.

	\begin{figure}
 		\centering		
 		\includegraphics[width=.85\linewidth, trim=0mm 0mm 0mm 0mm, clip]{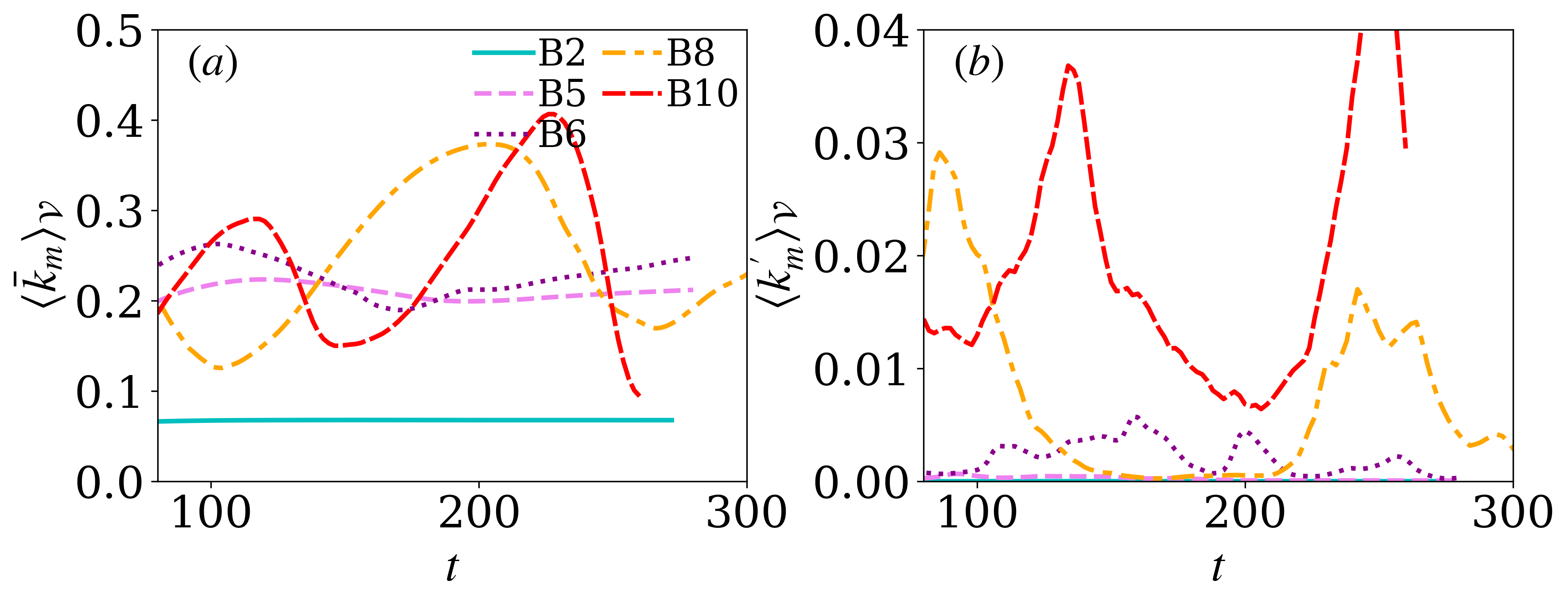}
 		\caption{Time series of volume-averaged (a) MKE and (b) TKE  for the five cases B2-B10 of figure~\ref{fig:ins_flow}. A snapshot of density and TKE in B6 at $t=160$ was shown in figure~\ref{fig:tke_window}(a-b).
 		}
 		\label{fig:tke_ts}
 	\end{figure}

	\Cref{fig:tke_ts} shows timeseries of the duct-volume-averaged MKE $\langle \bar{k}_{m} \rangle_\mathcal{V}$ (panel a) and TKE $\langle k^\prime_{m} \rangle_\mathcal{V}$ (panel b) for the five regimes shown in figure~\ref{fig:ins_flow}.
	The laminar (B2) and the stationary wave (B5) cases quickly reach a steady state  with constant or near-constant MKE (panel a)  and zero or near-zero TKE (panel b). We note that the flow in B5 is faster than in B2 as their MKE plateau at $\approx 0.2$ and $\approx 0.06$, respectively. The travelling wave (B6) case shows more fluctuations in the MKE (but also around $\approx 0.2$), and a larger TKE fluctuating between $\approx 0.001-0.005$. The intermittent  and turbulent cases (B8 and B10) show much larger fluctuations in MKE and TKE, and a significantly larger TKE than in the travelling wave case. In these two cases, the MKE and TKE fluctuations are of comparable magnitude to their temporal mean. The TKE fluctuations are particularly striking, showing that the flow, even when averaged over the entire duct volume, is alternating between phases of intense and weak turbulence.
	{    These fluctuations appear to be quasi-periodic with a period of $\approx 100$ advective time units, corresponding to approximately one and a half full-duct transit times at advective speed 1. Although long known from experiments, the mechanisms responsible for these fluctuations remain poorly understood and beyond the scope of this paper.}
	The intermittent regime (B8) differs from the turbulent regime (B10) in that its TKE occasionally drops to zero for extended periods of time (here $150\lesssim t \lesssim 210$); specifically, the flow relaminarises in the duct. This never happens in the turbulent regime, although it does feature cycles of weaker and stronger turbulence.

Our MKE and TKE timeseries are approximately similar to their salt-stratified experimental counterparts in \cite{lefauve_experimental2_2022} (see their figure 1(\textit{a,b})  and figure 3(\textit{f,i,o,r}), noting that they correspond to $\mathrm{Pr}\approx700$). This agreement between the DNS and experiments extends from the values of mean MKE $\approx 0.2$ in the W/I/T regimes, to the mean TKE $\approx 0.01-0.02$ in the I/T regimes, corresponding to a typical turbulent/mean velocity ratio of $\sqrt{k^\prime_m/\bar{k}_m} \approx 20-30$\,\% (although the ratio appears to be slightly higher in the DNS B10 than in the experiments T2 and T3). The large temporal fluctuations in B10  (which is at the limit of our computational resources) are typical of a flow near the I/T regime transition rather than well into the T regime, as was already clear from its location on the regime diagram (figure~\ref{fig:regime}(b)). The experimental time series of TKE (see \citet{lefauve_experimental2_2022}, figure 3(\textit{o,r}) in their datasets T1 and T3) suggest that these temporal fluctuations would significantly decrease in a more highly turbulent flow at higher $\mathrm{Re}$ and $\theta$ (i.e. that the TKE would become increasingly steadily sustained at a higher level). Our B8  has a time series similar to their dataset T1 (both being near the I/T transition), whereas their more turbulent dataset T3 is further away from the I/T transition. 

Our MKE and TKE timeseries are generally {out of phase} in time in the I and T regimes (figure~\ref{fig:tke_ts}), although the anti-correlation is less clear in the T regime. In other words, in B8, the MKE tends to decrease as the TKE increases (i.e. turbulence slows down the mean flow), and vice versa (the MKE tends to increase when the TKE decreases or is zero). 
This  behaviour, wherein the mean flow and the turbulence appear to regulate one another, supports the ideas of `self-organised criticality' and `marginal stability' discussed previously. In B10, this anticorrelation holds until $t\approx 200$, at which point the TKE increases rapidly while the MKE continues increasing, leading both TKE and MKE to peak approximately at the same time. In the T regime the mean flow thus appears closer to a turbulent threshold, such that perturbations grow more readily without `waiting' for the mean flow to fully accelerate. Equivalently, in the T regime (which has the highest $\theta \mathrm{Re}$), the mean flow is able to keep accelerating despite the growing turbulence, presumably due to a higher forcing (because it is proportional to $\theta$) and a lower TKE dissipation  than in the I regime (because it is inversely proportional to $\mathrm{Re}$) .

\begin{figure}
 		\centering		
 		\includegraphics[width=\linewidth, trim=4mm 0mm 0mm 0mm, clip]{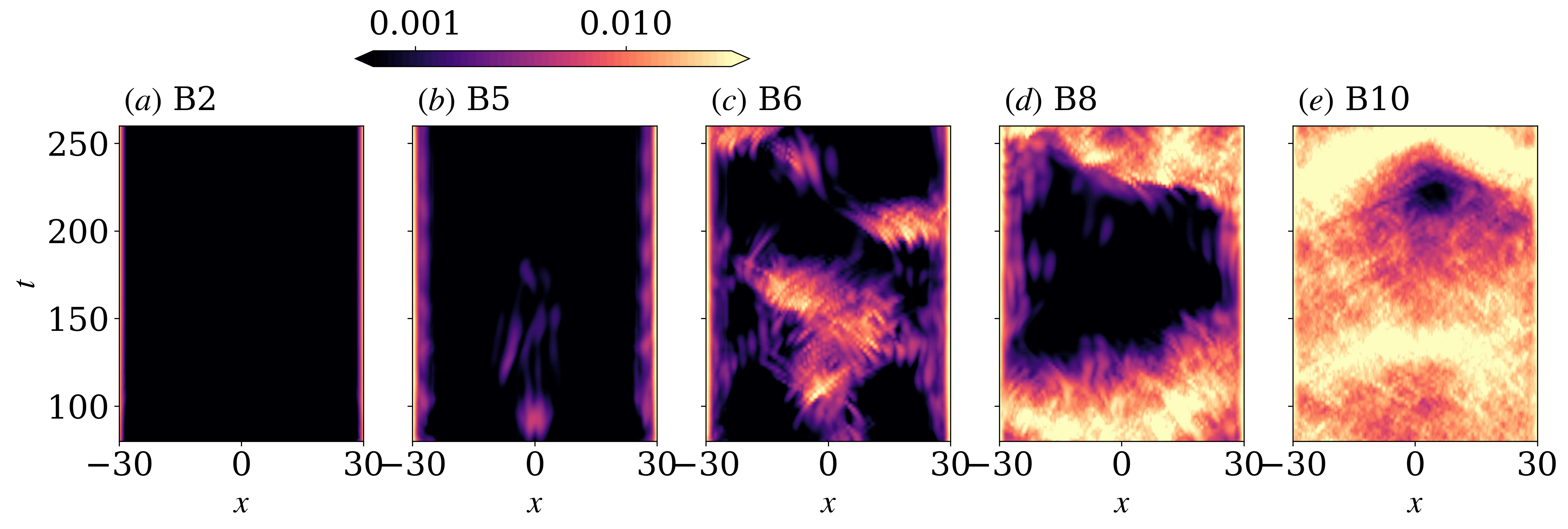}
 		\caption{Spatio-temporal diagram of TKE  $\langle k^\prime_m \rangle_{y,z}(x,t)$ for $t\in [80, 260]$ (after the initial transients) in DNS $(a)$ B2, $(b)$ B5, $(c)$ B6, $(d)$ B8, $(e)$ B10. 
 		 Note the colorbar is in log scale here. 
 		}
 		\label{fig:tke_xt}
 	\end{figure}

Finally, figure~\ref{fig:tke_xt} shows $x-t$ diagrams of TKE (averaged along $y$ and $z$) for B2 to B10 (left to right) after the initial transients ($t>80$) {have decayed}. In the L regime (B2,  figure~\ref{fig:tke_xt}(a)), the TKE is negligible except very near the ends of the duct $|x|\approx 30$, where tiny fluctuations are found where the exchange flow discharges into the reservoirs. In the stationary W regime (B5, panel b), the TKE at both ends is higher, extends a little further into the duct, and is also occasionally visible near the centre of the duct $|x|\approx 0$ where it appears to remain stationary. Similarly, the `end waves' do not propagate into the duct and are probably swept into the reservoirs, implying that their phase speed is smaller than the convective speed of the flow (i.e. that the flow may be critical or supercritical in these regions).
 	 	
    In the travelling W regime (B6; figure~\ref{fig:tke_xt}\textit{(c)}), larger TKE develops and it sometimes appears to propagate {along} the duct. These waves appear to be generated within the duct rather than travelling from the ends. In the I regime (B8, panel d), a laminar phase develops  between $120 \lesssim t \lesssim 220$, lasting over a full duct transit time (taking $\approx 2A=60$ ATU (advective time unit) at the maximum flow speed $\approx 1$). The boundary between laminar and turbulent phases appears to propagate from one end of the duct to the other. In the T regime, a quiescent patch develops near the centre of the duct just after $t=200$. The quiescent phase ends when energetic turbulent regions move in from both ends of the duct.
	
\subsection{Pressure}\label{sec:pressure}
	    
	We now analyse the pressure field throughout the interior of duct, which is inaccessible to experiments.

\begin{figure}
		\centering		
		\hspace{-0.6cm}
		\includegraphics[width=1.04\linewidth, trim=0mm 0mm 0mm 0mm, clip]{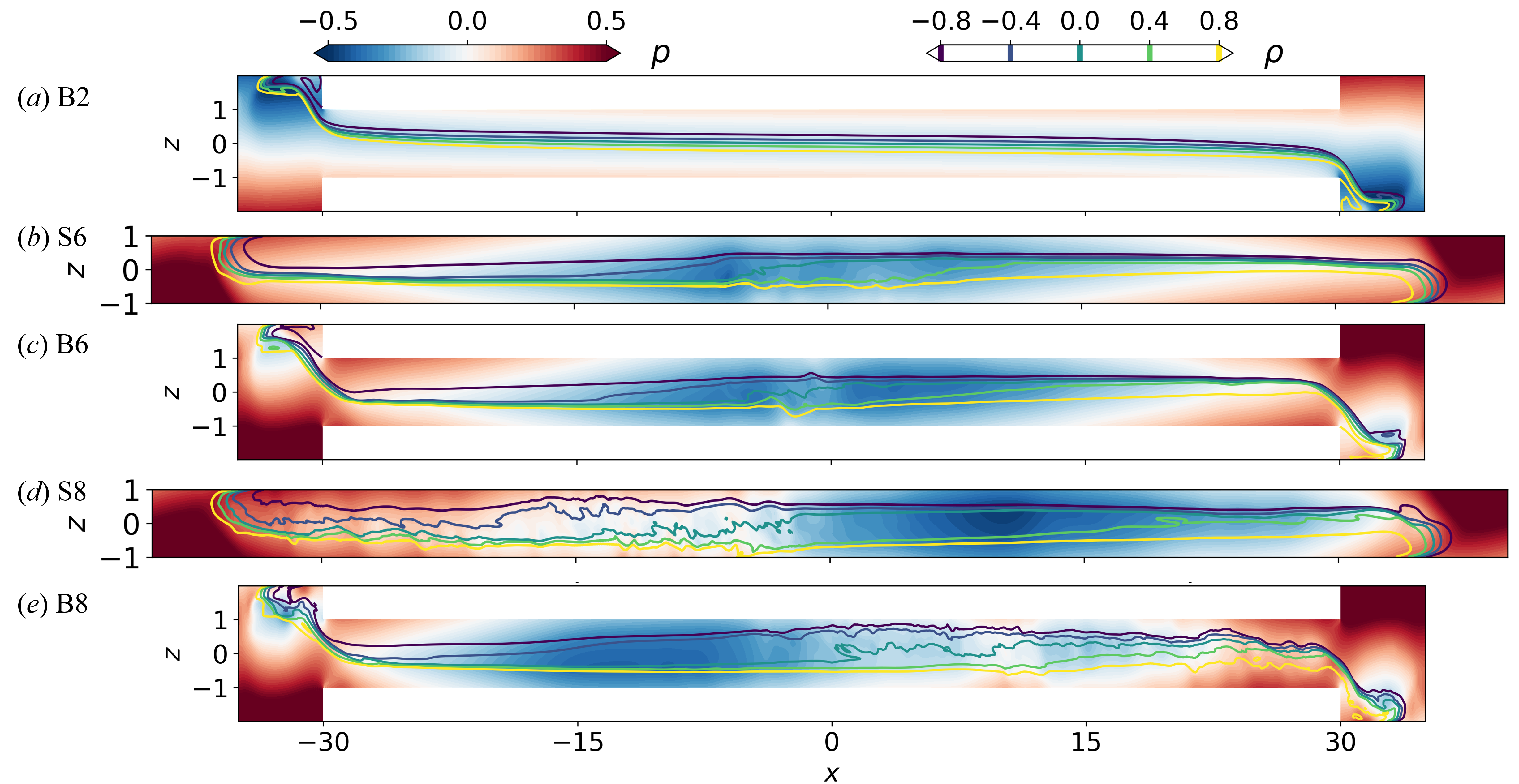}		
		\caption{{Spanwise-averaged pressure field snapshots (colours) superimposed with five isopycnals (lines) $\rho=0,\pm 0.4, \pm 0.8$ in DNS (a) B2, (b) S6, (c) B6, (d) S8, and (e) B8. } }\label{fig:inst_rho_p}
	\end{figure}

	\Cref{fig:inst_rho_p} shows representative snapshots of the spanwise-averaged pressure $\langle p \rangle_y$ and five equally spaced isopycnals for B2, S6, B6, S8, and B8. Note that our definition of the non-dimensional density $\rho$ as the perturbation around the reference $\rho_0$ (see \S\ref{sec:gov-eqs}) implicitly subtracts the hydrostatic pressure due to $\rho_0$ in the reservoirs. The pressure distribution is qualitatively similar in the W and I regimes (S6, B6, S8, B8; figure~\ref{fig:inst_rho_p}(b-e)), but  different in the L regime (B2, figure~\ref{fig:inst_rho_p}(a)).
	
	In the L flow inclined at $\theta=2^\circ$ (panel a), the pressure conforms to what we expect from an exchange flow in a horizontal duct where $\theta$ does not play a major role (see the sketch in \cite{lefauve2018waves}, figure 1.4). Essentially, each layer experiences a favourable streamwise pressure gradient: $-\p_x p > 0$ in the right-flowing lower layer where $u>0$, causing a convective acceleration along the duct, $u \p_x u >0$, and vice versa, $-\p_x p < 0$ in the left-flowing upper layer where $u<0$, causing the expected $u \p_x u <0$ ($\p_x u >0$ in both layers). This is achieved by a high-pressure zone in the bottom left and top right reservoirs (in red), and a low-pressure zone in the top left and bottom right reservoirs (in blue), as would be naturally obtained by the hydrostatic equilibrium of two solutions having different densities and required to match hydrostatic pressures at mid-height. 
	    
	However, in flows inclined at $\theta=6^\circ$ and $8^\circ$ (figure~\ref{fig:inst_rho_p}(b-e)), the pressure  has a large-scale global minimum in the centre of the duct (in blue). This low-pressure zone causes both layers to experience a favourable pressure gradient over approximately the first half of their course, causing the fluid to accelerate as it flows towards the centre of the duct, but an \textit{adverse} pressure gradient over the second half of their course, causing the fluid to decelerate from the centre of the duct as it flows away from the centre of the duct. This adverse pressure gradient confirms the predictions of \cite{lefauve_experimental1_2022} (see their \S~4.3) who, without having direct access to the pressure field, found that in most data sets the Reynolds-averaged budget {implied} the existence of an adverse pressure gradient.

	    Furthermore, these features of the pressure distribution are found in both the BR and SR {geometries} (compare figure~\ref{fig:inst_rho_p}(b,c) and figure~\ref{fig:inst_rho_p}(d,e)) as the outflowing layers must decelerate in both geometries. In the BR (and by extension in the AR and Bench.), this occurs since the streams encounter fluid at rest in a large reservoir; in the SR this occurs since the streams encounter the artificial forcing region where the flow is brought to rest. This suggests that the SR appears to adequately mimic the Bench. despite the absence of reservoirs, which may help reduce computational costs further in future studies.

    
     {Finally, we address the impact of the adverse pressure gradient on the density field and its interface(s). The isopycnals in figure~\ref{fig:inst_rho_p}(a-d) (the yellow lines denoting the lower, densest layer, and the dark blue lines denoting the upper, lightest layer) show that the central low-pressure zone is linked with an increasing depth of each layer along the direction of the flow. This is expected from mass conservation along a straight duct: an accelerating layer (because of a favourable pressure gradient or gravitational forcing) must become thinner along its course, and vice versa, a decelerating layer (caused by an adverse pressure gradient) must become thicker. We also see from the isopycnals that this thickening is associated with the emergence of displaced isopycnals (in B6, S6) and turbulence (in B8, S8). Note that this interface thickening implies the occurrence of internal hydraulic `jump' and the set-up of hydraulic control~\citep{meyer2014stratified,lefauve_buoyancy_2020} in the middle of the duct. This effect of internal hydraulics is beyond the scope of this paper and will be revisited in more detail in our future work.}

\subsection{{Turbulent energy fluxes }} \label{sec:turb_fluxes}

{We conclude this section with an analysis of the turbulent energy fluxes in datasets B5, B6, B8 and B10 and use it to demonstrate how DNS data allows us to overcome the current experimental limitations identified in \cite{lefauve_experimental2_2022} (their appendix B) and improve our physical understanding of stratified turbulent mixing. }

{To do so, we adopt the `shear-layer' non-dimensional framework of \cite{lefauve_experimental1_2022} (\S~3.3) by rescaling all velocities such that the $(x,t)$-averaged $\langle u \rangle_t(y=0,z)$ has extrema $\pm 1$, and rescaling all spatial variables such that the $z$ location of these extrema are located at $\pm 1$ (whereas previously the top and bottom walls were located at $\pm 1$); we call this central region of non-dimensional height $=2$ the shear-layer. This effectively rescales the effective Reynolds number and bulk Richardson number of the flow, which we now denote $Re^s$ and $Ri_b^s$ respectively, and allows for more meaningful comparison of datasets with one another as well as with the literature. Following \cite{lefauve_experimental1_2022} we further remove from our analysis all data outside the shear layer, i.e. exclude the top and bottom near-wall boundary layers in $z$, as well the boundary layers in $y$ (where the peak $|u|$ is less than 0.7), in order to focus on the  `core' region with turbulent activity.}

{We then define the non-dimensional time and volume-averaged mean kinetic energy $\bar{K} = \langle k_m \rangle$ and turbulent kinetic energy $K' = \langle k'_m \rangle$, as well as the mean scalar variance $\bar{K}_\rho \equiv Ri^b_s \langle \bar{\rho}_m^2 /2 \rangle $  and turbulent scalar variance $K_\rho' \equiv Ri^b_s \langle \rho'_m{}^2 /2 \rangle$. Note that the subscript $m$ denotes the moving average introduced in figure~\ref{sec:flowDyn}, and that the multiplying factor $Ri_b^s$  allows us to interpret $K_\rho,K_\rho'$ as proxies for potential energy under linear stratification. The simple bracket averaging  $\langle \cdot \rangle$ denotes a combined three-dimensional volume averaging over the shear layer region  and along the central two-thirds of the duct (by excluding one averaging window $\Delta L$ on either end) and time-averaging averaging over $t\in [80, 280]$ (focussing on the established dynamics as in figure~\ref{fig:tke_xt}).}

{Considering the evolution equations of these four energy reservoirs under a set of `safe' approximations in SID, \cite{lefauve_experimental2_2022} derived the following approximate balances between energy fluxes in a statistically steady state:
\begin{subeqnarray} \label{dKdt_0}
 \mathcal{P}  &\approx& \mathcal{F} -  \bar{\epsilon} \ \quad \text{(production of } K' = \text{forcing} - \text{laminar dissipation)} \slabel{dKdt_0-1}  \\ 
\mathcal{E}&\approx&  \mathcal{P} - \mathcal{B}   \ \   \ \ \text{(dissipation of } K' = \text{production} - \text{buoyancy flux)} \slabel{dKdt_0-2} \\ 
   \mathcal{P}_\rho  &\approx& \Phi^{\bar{K}_\rho}  \quad \ \   \text{(production of } K'_\rho = \text{boundary net flux of unmixed fluid)}  \slabel{dKdt_0-3} \\ 
  \chi  &\approx& \mathcal{P}_\rho \ \,  \qquad \text{(dissipation of } K'_\rho \text{ i.e. mixing} = \text{production)} \slabel{dKdt_0-4}
\end{subeqnarray}
where the eight fluxes are:
\begin{equation} \label{dKdt-terms-1}
 \mathcal{P} \equiv - \langle u'v'\p_y \bar{u}+ u'w' \p_z \bar{u} \rangle, \ \  \ \mathcal{F}  \equiv  Ri_b^s \, \sin \theta \, \langle \bar{u}\bar{\rho}\rangle ,  \ \ \ 
\bar{\epsilon} \equiv   \frac{2}{Re^s} \langle||\bar{\mathsf{\mathbf{s}}}||^2\rangle, \ \ \ \mathcal{E} \equiv  \frac{2}{Re^s} \langle ||\mathsf{\mathbf{s}}'||^2 \rangle
\end{equation}
\begin{equation} \label{dKdt-terms-2}
  \Phi^{\bar{K}_\rho}  \equiv  -Ri_b^s \Big\langle \frac{ \big( u\frac{\rho^2}{2}\big)\big|^{x^+}_{x^-}}{x^+ - x^-} \Big\rangle, \  \  \mathcal{B}   \equiv  Ri_b^s \langle w' \rho' \rangle, \  \ \ \mathcal{P}_\rho \equiv - Ri_b^s  \langle w' \rho' \p_z \bar{\rho} \rangle, \  \
  \ \chi \equiv \frac{Ri_b^s}{Re^s \, Pr}\langle |\boldsymbol{\nabla} \rho'|^2\rangle,
\end{equation}
where $\bar{\mathsf{\mathbf{s}}}, \mathsf{\mathbf{s'}}$ are the strain rate tensors of the mean and turbulent velocity fields, respectively. From  \eqref{dKdt-terms-1} and henceforth we omit the subscripts $m$, using the moving average  in all bar and prime quantities. 
Equations \eqref{dKdt_0-2} and \eqref{dKdt_0-4} are the classical balances of \cite{osborn_estimates_1980} and \cite{osborn_oceanic_1972} respectively, while  \eqref{dKdt_0-1}  \eqref{dKdt_0-3}   are specific to SID: the mean kinetic energy is sustained $\mathcal{F}$ by the gravitational acceleration and the mean scalar variance is sustained through $\Phi^{\bar{K}_\rho}$ by the inflow of unmixed fluids from the reservoirs into the volume of interest.}

\begin{figure}
		\centering		
		\includegraphics[width=1.0\linewidth, trim=0mm 0mm 0mm 0mm, clip]{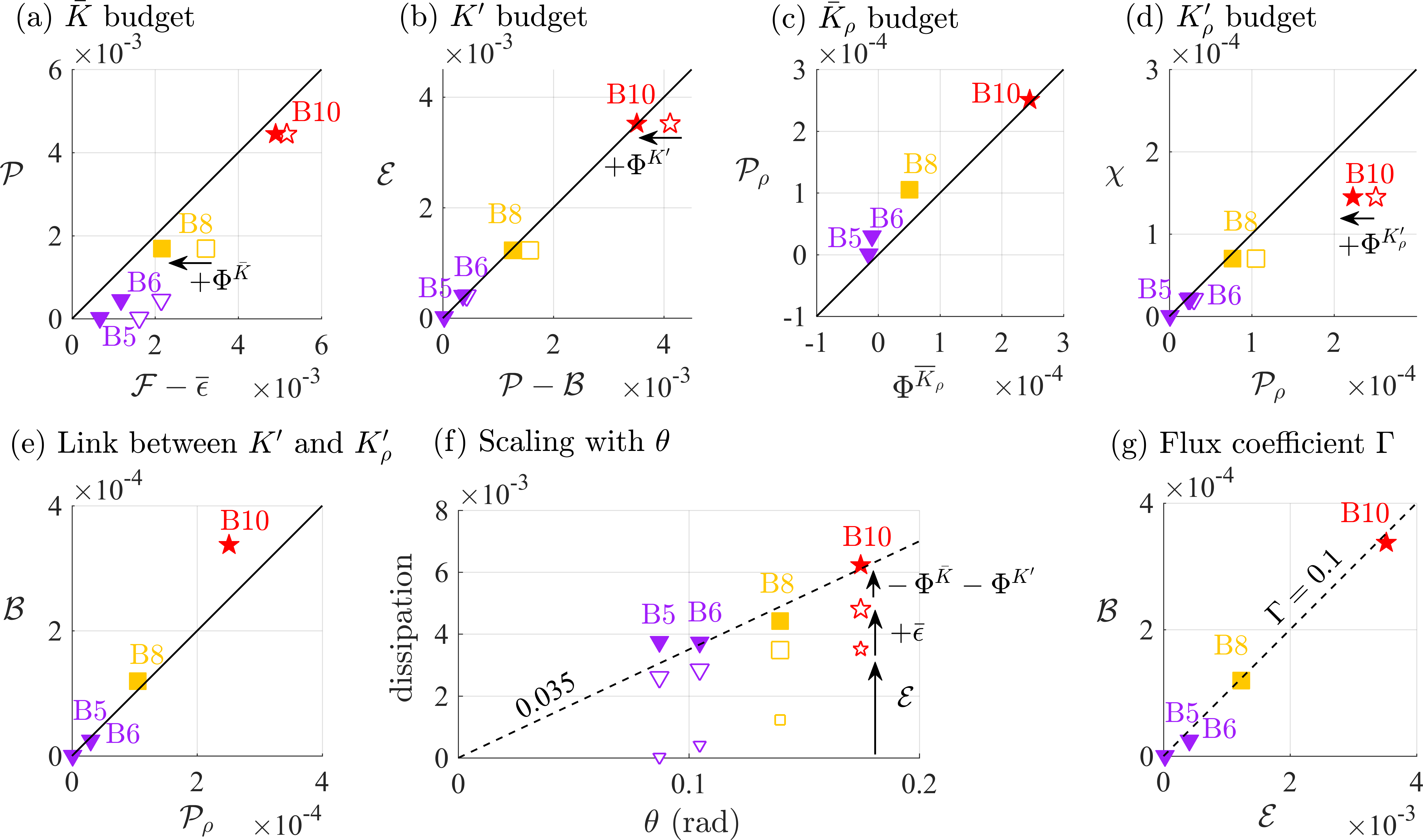}
		\caption{Correlation between the time- and volume-averaged energy fluxes \eqref{dKdt-terms-1}-\eqref{dKdt-terms-2} in B5, B6, B8, B10. (a-d) Verification of the energy balances in \eqref{dKdt_0} (LHS vs RHS), full symbols denoting the minor correction of boundary fluxes \eqref{bfluxes}, (e-f) Verification of the empirical relations \eqref{emp-rel}. (g) Determination of the empirical flux coefficient \eqref{Gamma} suggesting $\Gamma=0.1$. } \label{fig:turb_fluxes}
	\end{figure}

{\Cref{fig:turb_fluxes}(a-d) demonstrate how closely the four balances  \eqref{dKdt_0}(a-d) hold in DNS, where the thick diagonal solid line denotes equality between the left-hand side (LHS, vertical axis) and right-hand side (RHS, horizontal axis) of each equation. The  empty symbols in panels a,b,d denote the value of the LHS and RHS exactly as in \eqref{dKdt_0}(a,b,d), and are generally in balance, except in a few cases. In these few cases, the balance is improved by the solid symbols obtained after adding the following boundary fluxes
\begin{subeqnarray} \label{bfluxes}
\Phi^{\bar{K}} &\equiv&  - \Big\langle \frac{\big(u (\frac{\mathbf{u}^2}{2}+p)\big)\big|^{x^+}_{x^-}}{x^+ - x^-}   \Big\rangle  \  \text{to the RHS of } \eqref{dKdt_0-1}, \slabel{bflux1} \\
\nonumber \\
\Phi^{K'}  &\equiv&  - Ri_b^s \Big\langle \frac{\big(u' (\frac{\mathbf{u'}^2}{2}+p')\big)\big|^{x^+}_{x^-}}{x^+ - x^-}  +  \frac{\big(v' (\frac{\mathbf{u'}^2}{2}+p')\big)\big|^{y^+}_{y^-}}{y^+ - y^-} +   \frac{\big(w' (\frac{\mathbf{u'}^2}{2}+p')\big)\big|^{z^+}_{z^-}}{z^+ - z^-}  \Big\rangle \ \text{to the RHS of }  \eqref{dKdt_0-2} ,  \nonumber \\ 
\slabel{bflux2} \\
\Phi^{K'_\rho} &\equiv& - Ri_b^s \Big\langle \frac{\big(u'\frac{\rho' {}^2}{2})\big|^{x^+}_{x^-}}{x^+ - x^-}  +  \frac{\big(v' \frac{\rho'{}^2}{2}\big)\big|^{y^+}_{y^-}}{y^+ - y^-} +   \frac{\big(w' \frac{\rho'{}^2}{2}\big)\big|^{z^+}_{z^-}}{z^+ - z^-}  \Big\rangle  \  \text{to the RHS of } \eqref{dKdt_0-4},  \slabel{bflux3} 
\end{subeqnarray}
where $x^\pm, y^\pm, z^\pm$ are the edges of our shear-layer averaging domain, and we neglected in \eqref{bflux1} the spanwise and vertical mean transport, in  \eqref{bflux1}-\eqref{bflux2} the work of viscous forces, and in \eqref{bflux3} the transport by molecular diffusion, in order to focus on the dominant contributions. The fact that these fluxes improve the balance only slightly demonstrates that they can, to a reasonable approximation, be neglected in SID energetics. This  result was hypothesised in \cite{lefauve_regime_2019} and \cite{lefauve_experimental2_2022} but their experiments could not clearly confirm it due to excessive noise in the computation of \eqref{bfluxes} and the lack of the pressure. After correction, the only remaining discrepancy is found in $\chi$ in the most turbulent flow B10 (panel d)  B10, where  it is $35~\%$ below the expected value to balance $\mathcal{P}_\rho$. }

{Over a third of this discrepancy ($13~\%$ out of $35~\%$) is explained by our neglect of molecular diffusion $Ri_b^s/(Re^s Pr)|\boldsymbol{\nabla}\bar{\rho}|^2$ in \eqref{dKdt_0-4}, which at these values of $Re^s$ and $Pr$, is `only' a factor of five smaller than $\chi$. The remaining two thirds of this discrepancy do not appear to be due to under-resolution, since our spatial grid approaches the Batchelor length-scale computed in the shear-layer $[\Delta x,\, \Delta y,\, \Delta z] = [3.3, \,   2.3,  \, 2.3] \, \ell_B$, where  $\ell_B \equiv \langle \mathcal{E} \rangle^{-1/4}(Re^s)^{-3/4}Pr^{-1/2}= 0.01$ in non-dimensional shear-layer units. Furthermore, we verified that $\chi$ was already converged after comparing with a coarser grid (by a factor of $\approx 2$ in $x$ and $y$, see \cref{tab:simul_overview}). Closer evaluation of the underlying time series shows that $\chi$ undergoes two cycles over $t\in[80,280]$, much like the TKE in figure~\ref{fig:tke_ts}(b), with peaks being 10 times larger than the troughs. Therefore it is possible that our  insufficiently long time averaging window (containing only two extreme events) may yield a time-averaged $\chi$ slightly below what it would be when averaged over longer times.  }


%
%

{Having largely verified \eqref{dKdt_0}, we now examine in figure~\ref{fig:turb_fluxes}(e-f) two robust empirical relations from the SID experimental literature in very turbulent flows ($\mathcal{E} \gg  \bar{\epsilon}$):
\begin{subeqnarray} \label{emp-rel}
 \mathcal{P_\rho}  &\approx&  \mathcal{B}  \qquad \ \, \ \ \qquad \text{because } \p_z \langle \rho\rangle_{x,t} \approx -1 \text{ in the the shear layer} \slabel{emp-rel-1}\\
  \mathcal{E}  +  \bar{\epsilon}   &\approx&  \mathcal{E}   \approx 0.035 \theta  \ \ \ \ \text{because of hydraulic control and } Ri^b_s \approx 0.10-0.15 \slabel{emp-rel-2}
\end{subeqnarray}
where $\theta$ is in radians \citep{lefauve_experimental2_2022}. Our DNS data generally confirm \eqref{emp-rel-1} (panel e) and  \eqref{emp-rel-2} (panel f) but with two reservations. First, in B10, $\mathcal{P}_\rho$ is $25~\%$ below the expected value $\mathcal{B}$ as a result of the mean vertical density gradient being slightly weaker in our DNS (at $Pr=7$) than in the experiments (at $Pr=700$). Second, although the total dissipation $\mathcal{E}+\bar{\epsilon}$ corrected with the appropriate boundary fluxes (full symbols) follows \eqref{emp-rel-2}  (in particular since our $Ri_b^s$ indeed converges to $0.12-0.14$ in all datasets B5-10), the agreement is less good for the turbulent dissipation alone $\mathcal{E}$ (smaller empty symbols). In other words, our ability to fully capture $\mathcal{E}$ and all boundary fluxes in DNS allow us to quantify the relative importance of the (subdominant) terms $\Phi^K,\Phi^{K'}$ and $\bar{\epsilon}$ in SID energetics. We hypothesise that stronger turbulence at higher values of $\theta Re^s$ would see these subdominant terms plateau, and  $\mathcal{E}$ follow $0.035 \theta$ increasingly closely.}

{We now move to the ultimate goal of this energetics analysis, and more broadly of research into turbulent mixing, which is to eventually connect all turbulent fluxes to known non-dimensional parameters of the flow in a closed system of equations. In the asymptotic `strong SID turbulence' scenario under which $\bar{\epsilon}$ becomes subdominant, we have seven key turbulent fluxes in \eqref{dKdt-terms-1},\eqref{dKdt-terms-2} and six independent equations:  the four equations \eqref{dKdt_0} expressing conservation of energy, and the two robust empirical equations \eqref{emp-rel}, one of which crucially involves the single input parameter $\theta$. To close the system, we  require a seventh independent equation, which we choose to be the classical flux parameter in the ocean mixing literature:
\begin{equation}\label{Gamma}
\Gamma \equiv \frac{  \mathcal{B} }{\mathcal{E}}.  
\end{equation}
Combining these seven equations in matrix form, and inverting this linear system, we deduce all fluxes in closed form as:
\begin{equation} \label{all_fluxes_closed_form}
  \begin{bmatrix}
   1 & -1 & 0 & 0 & 0 & 0 & 0 \\ 
      0 & -1 & 1 & 1 & 0 & 0 & 0 \\
         0 & 0 & 0 & 0 & 1 & -1 & 0 \\
            0 & 0 & 0 & 0 & 1 & 0 & -1 \\
               0 & 0 & 0 & 1 & -1 & 0 & 0 \\
                0 & 0 & 1 & 0 & 0 & 0 & 0 \\
                0 & 0 & \Gamma & -1 & 0 & 0 & 0 \\
   \end{bmatrix}
     \begin{bmatrix}
   \mathcal{F} \\
   \mathcal{P} \\
   \mathcal{E} \\
  \mathcal{B} \\
   \mathcal{P}_\rho \\
  \Phi^{\bar{K}_\rho}  \\
  \chi
   \end{bmatrix} =
        \begin{bmatrix}
  0 \\
 0\\
  0 \\
 0\\
 0\\
   0.0035\theta \\
  0
   \end{bmatrix} \ \Longrightarrow    \ \begin{bmatrix}
   \mathcal{F} \\
   \mathcal{P} \\
   \mathcal{E} \\
  \mathcal{B} \\
   \mathcal{P}_\rho \\
  \Phi^{\bar{K}_\rho}  \\
  \chi
   \end{bmatrix} =     0.035\theta \begin{bmatrix}
   1+\Gamma \\
   1+\Gamma  \\
   1 \\
  \Gamma \\
 \Gamma \\
  \Gamma  \\
  \Gamma
   \end{bmatrix} ,
\end{equation}
This expression highlights the importance of knowing the value of $\Gamma$, and its potential dependence on any of the non-dimensional flow parameters such as $\theta, Re$ or $Pr$, as the keystone to turbulent mixing in SID. \Cref{fig:turb_fluxes}(g) shows   $\mathcal{B}$ vs $\mathcal{E}$ (both of which are fully resolved in our DNS) which strongly support the constant value of $\Gamma\approx 0.1$ in the two most turbulent datasets (0.097  and 0.096 in B8 and B10 respectively). The experimental data \cite{lefauve_experimental2_2022} also found $\Gamma\approx 0.1$ but their insufficient spatial resolution to fully capture $\mathcal{E}$ led them to conjecture a slightly lower value in the range $0.05-0.07$. Our DNS allow us to confirm that $\Gamma\approx 0.1$ is a robust estimate, at least for turbulence at $Pr=7$ in this narrow region of the $(\theta,Re)$ space. }

\section{Conclusions} \label{sec:conclu}
	
In this paper, we performed and interpreted DNS of stratified shear flows in a long rectangular duct connecting two reservoirs. The flow is continuously forced by gravity by a modest positive tilt angle $\theta=0-10^\circ$, and has Reynolds number $\mathrm{Re}=400-1250$, bulk Richardson number $\mathrm{Ri}=0.25$, and Prandtl number $\mathrm{Pr}=7$.  Our results are summarised as follows.

\subsection{An efficient numerical paradigm for SID}

In \S\ref{sec:method} we presented a new numerical set-up (figure~\ref{fig:geom}) designed to closely mimic the experimental set-up of the stratified inclined duct (SID). 
We introduced a new forcing term in the reservoirs (figure~\ref{fig:geom}) that allows the exchange flow to be sustained indefinitely with small reservoirs, thus focusing our computational resources on the flow of interest within the duct. We also implemented an immersed boundary method (figure~\ref{fig:ibm}) to enforce the boundary conditions on the duct walls that match the experiments, i.e. no-slip for velocity and no-flux for density.
  
In \S\ref{sec:valid} we validated this numerical {configuration}. First, we showed that our artificial forcing in the reservoirs was necessary to sustain the exchange flow by `refreshing' any finite-sized reservoirs  beyond the short time-scale over which they would otherwise fill with mixed fluid (figure~\ref{fig:U_ts_com}). Second, we showed that small reservoirs combined with the appropriate forcing were sufficient to reproduce the flow of a Bench. case having very large reservoirs and no forcing (figure~\ref{fig:stat_forc}).

In \S\ref{sec:transition} we described the {properties} of increasingly disorganised and turbulent flow regimes found by increasing $\mathrm{Re}$ and $\theta$ (figure~\ref{fig:ins_flow}). These regimes are similar to those found in experiments where the stratification is achieved by temperature (approximately matching our $\mathrm{Pr}=7$), which further validates the relevance and accuracy of our DNS to faithfully reproduce experimentally-realisable flows. These flow regimes are generally found in the same region of $\theta-\mathrm{Re}$ parameter space as the experiments (figure~\ref{fig:regime}), with very little difference between the larger reservoir (AR) and the smaller reservoir (BR). This agreement between DNS and experiments carries over to more detailed flow characteristics visualised by instantaneous shadowgraph snapshots (figure~\ref{fig:SG}) and spatio-temporal diagrams (figure~\ref{fig:SG_zt}). 


In \S\ref{sec:addedvalue} we studied quantitative DNS diagnostics that complement experimental diagnostics. We first investigated the vertical velocity and density profiles. The gradient Richardson number (figure~\ref{fig:Stat_case}) displayed the same turbulent `equilibrium' with nearly uniform $\mathrm{Ri}_g \approx 0.10-0.15$ across the shear layer as in the experiments, despite the difference in $\mathrm{Pr}$. We then moved to the mean and turbulent kinetic energies  (figure~\ref{fig:tke_window}), introducing a moving-average  (in $x$) definition suited to our DNS data, focussing on the intermittency of the turbulence (figure~\ref{fig:tke_ts}).
We  investigated the spatio-temporal behaviour of the TKE (figure~\ref{fig:tke_xt}), exploiting the fact that our DNS data is 
available along the full length of the duct. We contrasted stationary and travelling waves, described where waves originate from, and how turbulence or relaminarisation sometimes occur synchronously along the duct, and sometimes in `waves' propagating at the advective speed. 
Next, we investigated the pressure field (figure~\ref{fig:inst_rho_p}) and discovered
a large-scale low-pressure zone inside the duct in all non-laminar flows, which was previously conjectured with experimental data, but only proven with DNS data.  We showed that our ad hoc forcing, even in the smallest geometry SR, was sufficient to reproduce this behaviour observed with real reservoirs. This low-pressure zone 
creates a favourable pressure gradient in both layers over roughly the first half of their transit, allowing them to accelerate as they flow in, but, crucially, an adverse pressure gradient over roughly the second half of their transit, causing them to decelerate before they flow out. This result suggests a potentially new  mechanism for hydraulic control in exchange flows tilted at a favourable angle $\theta$, which requires further study. {Finally, we have largely confirmed the simplified model in \eqref{dKdt_0} for the steady-state kinetic and scalar energy fluxes in SID turbulence, as well as the empirical relations in \eqref{emp-rel} (figure~\ref{fig:turb_fluxes}) hypothesised from experimental data. This allowed us to express in \eqref{all_fluxes_closed_form} all seven fluxes fully characterising the time- and volume-averaged turbulent energetics and mixing in SID as functions of $\theta$ and the flux coefficient $\Gamma$. Our data suggest $\Gamma \approx 0.1$ in the most turbulent flows, a value lower than the classical value of $0.2$ used in most of the ocean mixing literature.
  }

\subsection{Outlook}

This paper introduced a computationally efficient way to simulate realistic shear-driven stratified turbulence over long time periods, which shows {excellent}  agreement with the experiments, {with} all non-dimensional parameters being matched (at $Pr=7$). 
We consider this comprehensive agreement between highly-nonlinear numerical and experimental fluid dynamics to be the major result of this paper, and a milestone in the study of SID and stratified turbulence. Furthermore, the numerics add considerable value to the experiments by providing accurate and highly-resolved data over the entire domain, and by allowing arbitrarily long integration times with the addition of forcing terms in the reservoirs.

There is significant scope to build on this study, by overcoming technical challenges.  
For example, improved experimental technology is needed to obtain more accurate, higher-resolution data in more highly turbulent flows at $\mathrm{Re}=O(10^3-10^4)$. Increased computational power is also needed to match such $\mathrm{Re}$ and to tackle the differences between temperature and salt stratification in the range $\mathrm{Pr}=O(10^2)$. Finally, studying the slow, quasi-periodic dynamics of intermittent turbulence requires large physical reservoirs and data acquisition as well as long integration times and costly simulations. Nevertheless, as experimental technology improve and computational power increases, we anticipate that they will be able to cover a much larger range in parameter space and answer questions previously inaccessible to theory, observations, experiments or simulations alone.

	\vspace{0.3cm}
	\noindent	\textbf{Acknowledgements.} We are grateful to Dr Xianyang Jiang and Dr Gaopan Kong for their help in carrying out the new shadowgraph experiments for figure~\ref{fig:SG} and figure~\ref{fig:SG_zt}. We thank Dr Ricardo Frantz for his help with the development of our DNS with Xcompact3d. 
	
	\vspace{0.3cm}
	
		\noindent	\textbf{Funding.} This work was supported by the European Research Council (ERC) under the European Union Horizon 2020 Research and Innovation Grant No 742480 ‘Stratified Turbulence And Mixing Processes’ (STAMP). Part of the DNS were run with resources from Compute/Calcul Canada. A.L. is supported by a Leverhulme Early Career Fellowship. For the purpose of open access, the authors have applied a Creative Commons Attribution (CC BY) licence to any Author Accepted Manuscript version arising from this submission.

	\vspace{0.3cm}
	\noindent \textbf{Declaration of interests.} The authors report no conflict of interest.
	
		\vspace{0.3cm}

	\clearpage
	\bibliographystyle{jfm}
	\bibliography{main,AL_references_2022_01}



\end{document}